\documentclass{aa}
\usepackage{graphicx}
\usepackage{txfonts}
\usepackage{dcolumn}
\usepackage{multirow}
\usepackage{subfigure}
\newcolumntype{d}{D{;}{\pm}{-1}}
\usepackage{color}

\begin{document}

   \title{CHEERS: The chemical evolution RGS sample}

   \author{J. de Plaa\inst{1} \and J. S. Kaastra\inst{1,2} \and N. Werner\inst{3,4,5} \and 
           C. Pinto\inst{7} \and P. Kosec\inst{6,7} \and  Y-Y. Zhang\inst{8} \and F. Mernier\inst{1,2} \and
           L. Lovisari\inst{8,9} \and H. Akamatsu\inst{1} \and G. Schellenberger\inst{8,9} \and
           F. Hofmann\inst{10} \and T. H. Reiprich\inst{8} \and A. Finoguenov\inst{11,10} \and J. Ahoranta\inst{11} \and 
           J.S. Sanders\inst{10} \and A.C. Fabian\inst{7} \and O. Pols\inst{12} \and 
           A. Simionescu\inst{13} \and Jacco Vink\inst{14} \and H. B\"ohringer\inst{10}
          }

   \institute{SRON Netherlands Institute for Space Research, 
              Sorbonnelaan 2, 3584 CA Utrecht, The Netherlands.\\
              \email{j.de.plaa@sron.nl}
         \and
              Leiden Observatory, Leiden University, PO Box 9513, 2300 RA, Leiden, The Netherlands.
         \and
              MTA-E\"{o}tv\"{o}s University Lend\"{u}let Hot Universe Research Group, P\'{a}zm\'{a}ny P\'{e}ter s\'{e}t\'{a}ny 1/A, Budapest, 1117, Hungary
         \and 
              Department of Theoretical Physics and Astrophysics, Faculty of Science, Masaryk University, Kotlarsk\'{a} 2, Brno, 611 37, Czech Republic
         \and 
              School of Science, Hiroshima University, 1-3-1 Kagamiyama, Higashi-Hiroshima 739-8526, Japan
         \and    
              Kavli Institute for Particle Astrophysics and Cosmology, Stanford University, 452 Lomita Mall, Stanford, CA 94305-4085, USA
         \and
              Institute of Astronomy, Madingley Road, Cambridge CB3 0HA, UK.
         \and 
              Argelander-Institut f\"ur Astronomie, Universit\"at Bonn, Auf dem H\"ugel 71, 53121 Bonn, Germany.
         \and
              Harvard-Smithsonian Center for Astrophysics, 60 Garden Street, Cambridge, MA 02138, USA.
         \and
              Max-Planck-Institut f\"ur extraterrestrische Physik, Giessenbachstrasse 1, D-85748 Garching, Germany. 
         \and
              Department of Physics, University of Helsinki, FI-00014 Helsinki, Finland.
         \and  
              Dept. of Astrophysics/IMAPP, Radboud University, P.O. Box 9010, 6500 GL Nijmegen, The Netherlands.
         \and
              Institute of Space and Astronautical Science (ISAS), JAXA, 3-1-1 Yoshinodai, Chuo-ku, Sagamihara, Kanagawa, 252-5210, Japan.
         \and     
              Anton Pannekoek Institute/GRAPPA, University of Amsterdam, PO Box 94249, 1090 GE, Amsterdam, The Netherlands.   
              }

   \date{}

 
  \abstract
   {The chemical yields of supernovae and the metal enrichment of the intra-cluster medium (ICM)
   are not well understood. The hot gas in clusters of galaxies has been enriched with metals 
   originating from billions of supernovae and provides a fair sample of large-scale 
   metal enrichment in the Universe. High-resolution X-ray spectra of clusters of galaxies provide 
   a unique way of measuring abundances in the hot intracluster
   medium (ICM). The abundance measurements can provide 
   constraints on the supernova explosion mechanism and the initial-mass function of the 
   stellar population. This paper introduces the CHEmical Enrichment RGS Sample (CHEERS), 
   which is a sample of 44 bright local giant ellipticals, groups, and clusters of galaxies observed with XMM-Newton.}
   {The CHEERS project aims to provide the most accurate set of cluster abundances measured in X-rays using this sample. 
   This paper focuses specifically on the abundance measurements of O and Fe using the reflection 
   grating spectrometer (RGS) on board XMM-Newton. We aim to thoroughly discuss the cluster to cluster 
   abundance variations and the robustness of the measurements.}
   {We have selected the CHEERS sample such that the oxygen abundance in each cluster is detected at a level of 
   at least 5$\sigma$ in the RGS. The dispersive nature of the RGS limits the sample to clusters with
   sharp surface brightness peaks. The deep exposures and the size of the sample allow us to quantify 
   the intrinsic scatter and the systematic uncertainties in the abundances using spectral modeling techniques.}
   {We report the oxygen and iron abundances as measured with RGS in the core regions of 
   all 44 clusters in the sample. We do not find a significant trend of O/Fe as a function of cluster 
   temperature, but we do find an intrinsic scatter in the O and Fe abundances from cluster to cluster.  
   The level of systematic uncertainties in the O/Fe ratio is estimated to be around 20$-$30\%, while 
   the systematic uncertainties in the absolute O and Fe abundances can be as high as 50\% in 
   extreme cases. Thanks to the high statistics of the observations, we were able to identify 
   and correct a systematic bias in the oxygen abundance determination that was due to an inaccuracy 
   in the spectral model.}
   {The lack of dependence of O/Fe on temperature suggests that the enrichment of the ICM does not
   depend on cluster mass and that most of the enrichment likely took place before the ICM was formed. 
   We find that the observed scatter in the O/Fe ratio is due to a combination of intrinsic
   scatter in the source and systematic uncertainties in the spectral fitting, which we are unable
   to separate. The astrophysical source of intrinsic scatter could be due to differences in active galactic nucleus 
   activity and ongoing star formation in the brightest cluster galaxy. The systematic scatter is due to uncertainties
   in the spatial line broadening, absorption column, multi-temperature structure, and the 
   thermal plasma models.}


   \maketitle
%

\section{Introduction}

Line cooling of chemical elements from C to Fe plays an 
important role in the formation of galaxies, stars, and planets. Most of the 
elements in the Universe today are thought to have formed in star bursts 
at $z\approx2-3$ \citep{hopkins2006,madau2014}.
The hot intracluster medium (ICM) in groups and clusters of galaxies 
is an excellent probe of this chemical evolution in the dense regions of 
the Universe. Metals are accumulated over very long times ($>5$~Gyr) in 
the cluster centers, and the total mass of metals in the hot
plasma in the core is about a factor of 2--6 higher than the metal 
mass locked up in the galaxies \citep{renzini2014}. The abundances measured in the plasma 
thus provide a ``fossil'' record of the integral yield of all the different stars 
(releasing metals in supernova explosions and winds) that have left 
their specific abundance patterns in the gas before and during cluster evolution 
\citep[see, e.g.,][for a review]{deplaa2013,werner2008}.

\begin{figure}[t]
\includegraphics[width=1.0\columnwidth]{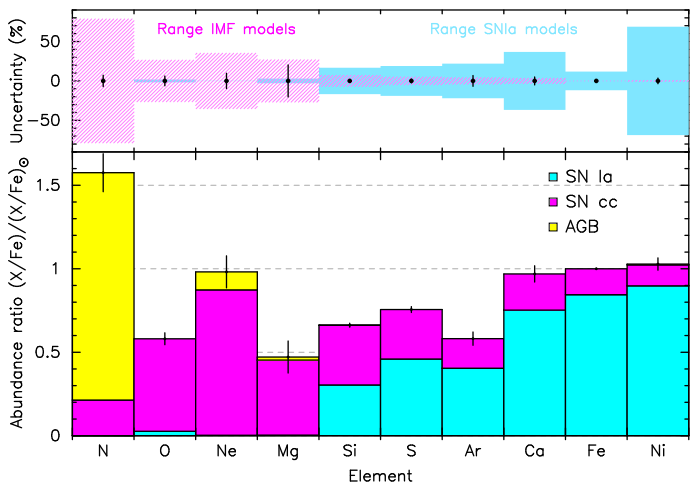}
\caption{Expected abundances measured in a typical long XMM-Newton observation of 120 ks 
(bottom panel). The estimates for the SNIa, SNcc, and AGB contribution are based on a 
sample of 22 clusters \citep{deplaa2007} and two elliptical galaxies 
\citep{grange2011}. The top panel shows the typical range in SNIa and IMF models with 
respect to the statistical error bars in the observation. Figure adapted from \citet{deplaa2013}.}
\label{fig:barplot}
\end{figure}

X-ray spectroscopy provides a precise measure of metal abundances in
the ICM. The observations with the European Photon Imaging Camera 
\citep[EPIC,][]{struder2001,turner2001} provide highly significant measurements of
the abundances of Si, S, Ar, Ca, Fe, and Ni. The
high-resolution X-ray spectra obtained with the Reflection Grating Spectrometer 
\citep[RGS,][]{herder2001} on board XMM-Newton, which resolves the
Fe-L complex into individual lines, allow for precise abundance measurements of
O, Ne, Mg, and Fe. In cooler clusters ($\lesssim$3 keV), RGS also detects lines from N 
\citep{sanders2011b}. Because non-equilibrium ionization and optical depth effects 
in the ICM are very weak, these abundances are more reliable than abundances measured in
stars or in the cold low-ionization interstellar medium. However, a thorough study
of systematic uncertainties in abundance measurements with RGS has not been performed 
to date. The error bars in Fig.~\ref{fig:barplot} show the expected statistical error bars 
on cluster abundances based on previous studies \citep{deplaa2007,deplaa2013}.

Most of the elements detected with XMM-Newton are produced by supernovae. 
Core-collapse supernovae (SNcc) produce large amounts of O, Ne, and Mg 
\citep[e.g.,][]{woosley1995,nomoto2006}, while type Ia supernovae (SNIa) produce large
quantities of Fe, Ni, and relatively little O, Ne, and Mg \citep[e.g.,][]{iwamoto1999,bravo2012}. 
The Si-group elements (Si, S, Ar, and Ca) are produced by both supernova types 
(see Fig.~\ref{fig:barplot}). N is produced mainly by asymptotic
giant branch (AGB) 
stars \citep{karakas2010,werner2006b,grange2011} and by winds of massive stars, especially in 
rotating stars and at low metallicity \citep[e.g.,][]{romano2010}. In this paper, we focus mainly on the O/Fe 
abundance ratio. The O/Fe, Ne/Fe, and Mg/Fe ratios are good indicators for
the relative contribution of SNIa with respect to SNcc. The knowledge of these ratios is 
important for determining the amount of Si-group elements produced by SNIa. 

SNIa have likely produced a substantial fraction of the Fe, Ni, and 
Si-group elements observed in cluster cores. The statistical precision of the abundance ratios 
derived from X-ray observations of nearby clusters and groups of galaxies in 
a typical XMM-Newton orbit of 120 ks is typically better than 10--20\%, while the 
spread in yields (see Fig~\ref{fig:barplot}, top panel) obtained from simulations 
assuming different SNIa explosion mechanisms can be up to a factor of a few for 
elements such as Ca and Ni \citep[e.g.,][]{iwamoto1999,badenes2003}. Accurate cluster abundances
therefore allow us to constrain supernova models.

Much of the uncertainty in SNIa yields is due to the variety in possible type Ia supernova
progenitors and the subsequent explosion mechanism. In recent years, the search for type Ia 
supernovae in galaxies has become much more efficient. Large samples of SNIa observed in mainly 
optical, infrared, and UV wavelengths revealed variations in SNIa properties that appear to 
correlate with the properties of the host stellar populations \citep[see, e.g.,][for a review]{howell2011,wang2012}.
In the single-degenerate (SD) supernova scenario, a carbon-oxygen white dwarf accretes matter from a 
non-degenerate companion star before it reaches the critical temperature for explosive
carbon ignition. It has become clear that the properties of the companion star are 
important for the properties of the type Ia explosion that follows after the accretion phase
and is one of the origins of the variety of SNIa that is observed. In addition, in the double-degenerate (DD) scenario, two white dwarfs merge and disintegrate in a
supernova explosion, creating yet another variety of type Ia supernovae. 

In addition to our lack of knowledge about the explosion mechanism, it is also unclear how the progenitor systems form.    
Attempts have been made to explain the observed type Ia rate theoretically through
simulations of the evolution of binary populations \citep[e.g.,][]{claeys2014}. 
This study showed that if the SNIa rate is due to the standard SD channel, the SNIa rate can 
be explained only under the assumption that the accretion onto the white dwarf is not limited 
(e.g., that the Eddington limit does not hold). 
The result of this and similar studies makes clear that type Ia supernovae are still a poorly 
understood phenomenon. Abundance ratios determined from clusters are therefore a key test
for binary population synthesis and SNIa explosion models. 

A simple test has been performed by, for example, \citet{deplaa2007}
using a sample of 22 clusters observed with XMM-Newton.
The authors analyzed the abundances of Si, S, Ar, Ca, Fe, and Ni within a radius
of 0.2R$_{500}$ from the cluster center. A good fit was obtained with a 
one-dimensional delayed-detonation model from \citet{badenes2003}, while 
models from \citet{iwamoto1999} were unsuccessful because they underestimated 
the Ca abundance. The model from \citet{badenes2003} that fitted the cluster
abundances in \citet{deplaa2007} also fits the abundances of the Tycho
supernova remnant \citep{badenes2006}, which is thought to have been a fairly typical
SNIa with an average luminosity. Recently, \citet{mulchaey2014} suggested that
a subclass of supernovae, called Ca-rich gap transients, may provide enough 
calcium to explain the high calcium abundance found in the ICM of clusters.  

However, the work of \citet{deplaa2007} only used abundances of elements heavier 
than Si determined from EPIC, and their O, Ne, and Mg measurements were determined 
from only two clusters for which they analyzed deep RGS spectra \citep{deplaa2006,werner2006}. In order to 
distinguish the SNcc and SN Ia contribution and place stronger constraints
on the SNIa explosion mechanism, accurate knowledge of the abundances of these elements 
from a large number of clusters is necessary.  Accurate measurements of O, Ne, and 
Mg, which are primarily products of SNcc, require the unique capabilities of the RGS. 

The O, Ne, and Mg yields from SNcc and N from AGB stars depend 
strongly on the progenitor mass. The difference in the total population yields between 
a top-heavy or Salpeter initial mass function (IMF) 
corresponds to a spread of 80\% in the N abundance, and a 30-40\% spread in O, Ne, and Mg 
(see Fig.~\ref{fig:barplot}). Unfortunately, N is not only produced in AGB stars, but also ejected 
in winds of massive stars before the SNcc explosion, especially in rotating stars and at low metallicity, 
which makes the models for the origin of N very uncertain \citep[see, e.g.,][]{romano2010}. Since 
the accuracy of the measured abundances is higher, a 
large sample of clusters with a broad range of masses, cool-core properties, and optical
characteristics of the central dominant (cD) galaxy may provide constraints on these models.
Ultimately, if we are also able to measure carbon and sodium, the abundance sets may provide 
a test of the IMF universality as well.

This paper introduces the CHEmical Enrichment RGS Sample (CHEERS) project, which mainly aims to 
obtain reliable chemical abundances in the intracluster medium of galaxy clusters through
deep XMM-Newton observations of 44 clusters. We introduce the sample and describe the selection of the clusters, 
which is optimized to exploit the RGS to the best of its abilities. The observations required to 
complete this sample were performed in AO-12 as part 
of an XMM-Newton Very Large Program. This paper reports the abundance results for oxygen and iron 
obtained from the RGS spectra. Because of the high spectral resolution in the soft X-ray band, the RGS is better 
capable of resolving the oxygen lines than EPIC. We aim to provide a thorough discussion about the 
reliability of the measurements and the robustness of the cluster to cluster variations.
This sample also provides very high quality XMM-Newton EPIC data. In two 
companion papers, we describe the EPIC abundance measurements of the other common elements using 
this sample \citep{mernier2016a} and the interpretation of the combined RGS and EPIC 
abundances \citep{mernier2016b}. The radial abundance profiles are studied in \citet{mernier2017}. 
In another companion paper, we report detections of 
nitrogen in a subset of the RGS observations \citep[][submitted]{mao2017}. Our measured abundances 
are relative to the proto-solar abundances by \citet{lodders2009} unless stated otherwise.
Error bars are given at the 1$\sigma$ (68\%) confidence level.

\section{Sample selection}

In order to study the chemical enrichment history of individual clusters of galaxies 
and the differences in enrichment between clusters, we need a 
moderately large sample of clusters with deep exposure times per cluster.
Because the sample of \citet{deplaa2007} lacked sufficient RGS coverage, 
we need to expand this sample to be able to divide 
it into subsamples of different cooling properties and study the 
spatial distribution of the elements. Our aim is to have a 'complete' sample of high-quality
RGS cluster spectra that can be obtained within a reasonable exposure time of $\lesssim200$ ks each. 
With 'complete', we mean that we aim to have observations of all suitable RGS cluster targets 
within a redshift of $z=0.1$. Obviously, many clusters already have deep RGS spectra, but the
XMM-Newton archive did not contain deep observations of all the suitable targets. We obtained 
deep XMM-Newton observations of 11 clusters in AO-12 as part of a very large program to 
complete the sample.

\begin{table*}
\caption{XMM-\textit{Newton} observations that define the complete CHEERS sample.}  
\label{tab:sample}      
\renewcommand{\arraystretch}{1.2}
\small\addtolength{\tabcolsep}{+2pt}
 
\begin{tabular}{c c c c c c c }     
\hline\hline            
Source                               &  ID $^{(a)}$          & Total clean time (ks) $^{(b)}$ & $kT$ (keV)  & $z^{(c)}$  & $N_{\rm H}$ ($10^{24}\,{\rm m}^{-2}$) $^{(d)}$\\  
\hline   
\multirow{1}{*}{\object{2A0335+096}} & 0109870101/0201 0147800201               &  120.5      &    3.0      &  0.0349    & 24.7  \\
\multirow{1}{*}{\object{A 85}}       &  \textbf{0723802101/2201}                &  195.8      &    6.1      &  0.0556    & 2.28  \\
\multirow{1}{*}{\object{A 133}}      &  0144310101 \textbf{0723801301/2001}     &  168.1      &    3.8      &  0.0569    & 1.09  \\
\multirow{1}{*}{\object{A 189}}      &  0109860101                              &   34.7      &    1.3      &  0.0320    & 4.31  \\
\multirow{1}{*}{\object{A 262}}      &  0109980101/0601 0504780101/0201         &  172.6      &    2.2      &  0.0161    & 5.76  \\
\multirow{1}{*}{\object{A 496}}      &  0135120201/0801 0506260301/0401         &  141.2      &    4.1      &  0.0328    & 5.37  \\
\multirow{1}{*}{\object{A 1795}}     & 0097820101                               &   37.8      &    6.0      &  0.0616    & 0.69  \\
\multirow{1}{*}{\object{A 1991}}     & 0145020101                               &   41.6      &    2.7      &  0.0586    & 1.96  \\
\multirow{1}{*}{\object{A 2029}}     & 0111270201 0551780201/0301/0401/0501     &  155.0      &    8.7      &  0.0767    & 2.75  \\
\multirow{1}{*}{\object{A 2052}}     & 0109920101 0401520301/0501/0601/0801     &  104.3      &    3.0      &  0.0348    & 2.21  \\
                                     & 0401520901/1101/1201/1301/1601/1701      &             &             &            &   \\
\multirow{1}{*}{\object{A 2199}}     & 0008030201/0301/0601 \textbf{0723801101/1201} &  129.7 &    4.1      &  0.0302    & 0.39  \\
\multirow{1}{*}{\object{A 2597}}     & 0108460201 0147330101 \textbf{0723801601/1701} & 163.9 &    3.6      &  0.0852    & 1.98  \\
\multirow{1}{*}{\object{A 2626}}     & 0083150201 0148310101                    &   56.4      &    3.1      &  0.0573    & 3.62  \\
\multirow{1}{*}{\object{A 3112}}     & 0105660101 0603050101/0201               &  173.2      &    4.7      &  0.0750    & 0.83  \\
\multirow{1}{*}{\object{A 3526}}     &  0046340101 0406200101                   &  152.8      &    3.7      &  0.0103    & 8.43  \\
\multirow{1}{*}{\object{A 3581}}     &  0205990101 0504780301/0401              &  123.8      &    1.8      &  0.0214    & 3.86  \\
\multirow{1}{*}{\object{A 4038}}     & 0204460101 \textbf{0723800801}           &   82.7      &    3.2      &  0.0283    & 1.03  \\
\multirow{1}{*}{\object{A 4059}}     & 0109950101/0201 \textbf{0723800901/1001} &  208.2      &    4.1      &  0.0460    & 0.71  \\
\multirow{1}{*}{\object{AS 1101}}    & 0147800101 0123900101                    &  131.2      &    3.0      &  0.0580    & 0.64  \\
\multirow{1}{*}{\object{AWM 7}}      & 0135950301 0605540101                    &  158.7      &    3.3      &  0.0172    & 9.20  \\
\multirow{1}{*}{\object{EXO 0422}}   & 0300210401                               &   41.1      &    3.0      &  0.0390    & 11.4  \\
\multirow{1}{*}{\object{Fornax}}     & 0012830101 0400620101                    &  123.9      &    1.2      &  0.0046    & 2.56  \\
\multirow{1}{*}{\object{HCG 62}}     & 0112270701 0504780501 0504780601         &  164.6      &    1.1      &  0.0140    & 4.81  \\
\multirow{1}{*}{\object{Hydra-A}}    & 0109980301 0504260101                    &  110.4      &    3.8      &  0.0538    & 4.18  \\
\multirow{1}{*}{\object{M 49}}       & 0200130101                               &   81.4      &    1.0      &  0.0044    & 2.63  \\
\multirow{1}{*}{\object{M 86}}       & 0108260201                               &   63.5      &    0.7      &  -0.0009   & 3.98  \\
\multirow{1}{*}{\object{M 87}}       & 0114120101 0200920101                    &  129.0      &    1.7      &  0.0042    & 1.44  \\
\multirow{1}{*}{\object{M 89}}       & 0141570101                               &   29.1      &    0.6      &  0.0009    & 2.12  \\
\multirow{1}{*}{\object{MKW 3s}}     & 0109930101 \textbf{0723801501}           &  145.6      &    3.5      &  0.0450    & 2.18  \\
\multirow{1}{*}{\object{MKW 4}}      & 0093060101 \textbf{0723800601/0701}      &  110.3      &    1.7      &  0.0200    & 1.25  \\
\multirow{1}{*}{\object{NGC 507}}    &  \textbf{0723800301}                     &   94.5      &    1.3      &  0.0165    & 7.33  \\
\multirow{1}{*}{\object{NGC 1316}}   & 0302780101 0502070201                    &  165.9      &    0.6      &  0.0059    & 1.90  \\
\multirow{1}{*}{\object{NGC 1404}}   & 0304940101                               &   29.2      &    0.6      &  0.0065    & 1.57  \\
\multirow{1}{*}{\object{NGC 1550}}   & 0152150101 \textbf{0723800401/0501}      &  173.4      &    1.4      &  0.0123    & 11.9  \\
\multirow{1}{*}{\object{NGC 3411}}   & 0146510301                               &   27.1      &    0.8      &  0.0152    & 4.25  \\
\multirow{1}{*}{\object{NGC 4261}}   & 0056340101 0502120101                    &  134.9      &    0.7      &  0.0073    & 2.86  \\
\multirow{1}{*}{\object{NGC 4325}}   & 0108860101                               &   21.5      &    1.0      &  0.0259    & 3.54  \\
\multirow{1}{*}{\object{NGC 4374}}   & 0673310101                               &   91.5      &    0.6      &  0.0034    & 3.38  \\
\multirow{1}{*}{\object{NGC 4636}}   & 0111190101/0201/0501/0701                &  102.5      &    0.8      &  0.0037    & 1.40  \\
\multirow{1}{*}{\object{NGC 4649}}   & 0021540201 0502160101                    &  129.8      &    0.8      &  0.0037    & 2.23  \\
\multirow{1}{*}{\object{NGC 5044}}   & 0037950101 0554680101                    &  127.1      &    1.1      &  0.0090    & 7.24  \\
\multirow{1}{*}{\object{NGC 5813}}   & 0302460101 0554680201/0301/0401          &  146.8      &    0.5      &  0.0064    & 3.87  \\
\multirow{1}{*}{\object{NGC 5846}}   & 0021540101/0501 \textbf{0723800101/0201} &  194.9      &    0.8      &  0.0061    & 4.26  \\
\multirow{1}{*}{\object{Perseus}}    & 0085110101/0201 0305780101               &  162.8      &    6.8      &  0.0183    & 20.0$^{(e)}$  \\
\hline                
\end{tabular}

$^{(a)}$ Exposure ID number. $^{(b)}$ RGS net exposure time. 
$^{(c)}$ Redshifts and temperatures are adapted from \cite{chen2007} and \cite{Snowden2008}. 
$^{(d)}$ Hydrogen column density determined using EPIC \citep{mernier2016a}.
$^{(e)}$ Hydrogen column determined from RGS observation (see Section~\ref{sec:nh-bias}).
New observations from our proposal are shown in boldface.\\
\end{table*}

\subsection{Selection of the proposed targets}

A substantial sample of suitable clusters is needed to study chemical enrichment
in different cluster environments. O, Ne, Mg, Ca and Ni in particular
are key elements for constraining the SNIa/SNcc contributions and the SNIa explosion 
mechanism. The nitrogen abundance, which is sensitive to the IMF, can then be 
measured in the subsamples that contain cool clusters. We need the RGS to measure 
the O and N abundance accurately. However, the varying spatial extent 
and brightness of clusters means that not all clusters are suitable RGS targets because 
the spatial surface brightness distribution of a cluster determines the spectral 
line width in the RGS (see Sect.~\ref{sec:rgs_broadening}). The clusters need to be 
bright and centrally peaked to resolve at least the brightest spectral lines. 
To select the brightest and the best-suitable clusters for the
RGS, we selected this 
sample mostly from the HIFLUGCS sample \citep{reiprich2002}. 

The SNIa/SNcc contribution ratio is mostly sensitive to the ratio between O and Fe, since they 
are the best-determined elements. Therefore, we required a 
statistical significance of about 10$\sigma$ on O in a single observation. With this 
criterion, we expect to obtain significances for Ne and Mg of 6$\sigma$ and 4$\sigma$ 
with the RGS, respectively, and $\sim 7\sigma$ for N in cool systems (kT$\lesssim$1 keV). 
This would in principle be enough to constrain SNcc and AGB models, for which N/Fe, O/Fe, 
Ne/Fe, and Mg/Fe vary between 30--80\% (see Fig~\ref{fig:barplot}). A 10$\sigma$ 
signal-to-noise ratio for O is a reasonable requirement for the selection of proposed clusters. 

A second criterion is to measure accurate abundances of less abundant 
elements. A key element is calcium \citep{deplaa2007}. 
An accurate Ca abundance guarantees even more 
precise values for the other elements.  The Ca/Fe ratio for different type Ia
models varies between 0.33--0.97 times solar \citep{werner2008}, so an accuracy
of 10\% solar on the Ca abundance would in principle be sufficient to
distinguish between different SNIa models or at least rule out certain models. 

Our final criterion for selecting the proposed clusters in AO-12 is an expected uncertainty of 
0.036 (10$\sigma$) solar for oxygen with the RGS and 0.10 times solar for Ca with EPIC. Clusters
were selected to be proposed if this criterion could be reached within an exposure time 
of $\sim$ 200 ks. In 85\% of all cases, oxygen gives the most stringent selection. The exposure times 
were increased by 40\% to account for possible loss of data due to 
soft-proton flares. The observations that were performed 
based on this selection are marked in boldface in Table~\ref{tab:sample}.

\subsection{Selection of archival observations}

In addition to the proposed targets that were observed by XMM-Newton in AO-12, we also 
considered archival cluster observations with high-quality RGS data. \citet{sanders2011}, 
for example, presented a list of high-quality RGS cluster observations. We reprocessed these
data, and in contrast to our proposed clusters, selected the clusters for which the oxygen 
abundance is detected at the 5$\sigma$ level to ensure that we obtained a reasonably 
large sample with sufficient spectral quality. Since we obviously do not have control over
the exposure time of archival data, we selected the clusters based on the bare minimum 
statistical quality data that we required. The final list of cluster observations is shown in Table~\ref{tab:sample}. 

\subsection{Other applications of the selected sample}

In this paper, we focus on the O/Fe abundance as measured with the RGS. Recently, the 
CHEERS sample also yielded science results other than abundance measurements. Some examples are the 
discovery of cool ($\sim$0.2 keV) gas in the CHEERS RGS spectra of elliptical galaxies 
\citep{pinto2014} and constraints on turbulent velocities measured with RGS using line 
broadening \citep[e.g.,][]{pinto2015}. The RGS data have also shown for \object{NGC 4636} 
that spatially resolved resonant scattering analysis is capable of revealing velocity 
structure in the ICM \citep{ahoranta2016}. This technique will soon be applied to 
more members of the sample in follow-up papers.

\section{Data analysis}

We used both the archival and new XMM-Newton exposures listed in Table~\ref{tab:sample}.
The observations were processed with the XMM-Newton Science Analysis Software
(SAS) version 14.0.0. For each observation, we extracted the event files from the ODF data files 
using calibration (CCF) files available on 2016/01/31. We used high-resolution spectra 
from the RGS and data from the MOS1 instrument to extract the spatial line profile used in the 
spectral fit of the RGS data. 

\subsection{RGS spectral extraction}

We processed the RGS data with the SAS task \textit{rgsproc} following the standard procedures.
In order to decrease the contamination from the soft-proton flares, we extracted
RGS light curves from CCD number\,9, where hardly any source emission is expected. 
We binned the light curves in 100\,s intervals and fit a Poissonian distribution 
to the count-rate histogram. We rejected all the time bins for which the number of counts lies 
outside the interval $\mu\pm2\sigma$, where $\mu$ is the fitted average of the distribution.
We used the resulting good time intervals (GTI) files to obtain the filtered event files. 

\begin{figure}
  \begin{center}
      \subfigure{ 
      \includegraphics[bb=15 15 515 348, width=9cm]{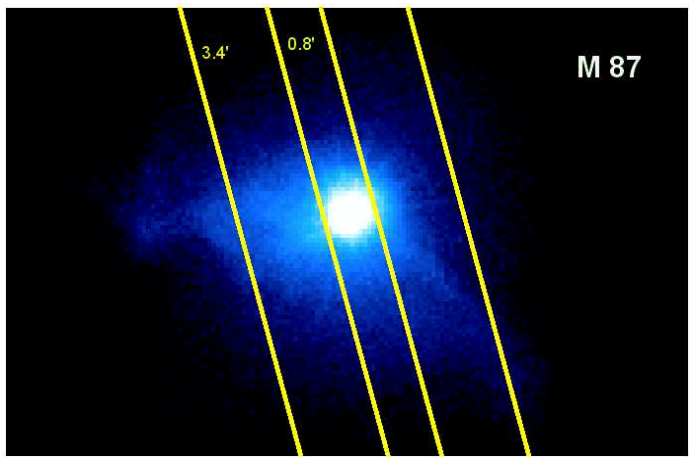}}
      \caption{RGS extraction regions and MOS\,1 stacked image of M\,87.}
          \label{fig:rgs_regions}
  \end{center}
\end{figure}

We extracted the RGS source spectra in a region centered on the peak of the source emission
with a width of 0.8$^{\prime}$. We used the model background spectrum created by the 
standard RGS pipeline, which is a template background file, based on the count rate in CCD\,9 of the RGS. 
In Fig.~\ref{fig:rgs_regions} we show the 0.8$^{\prime}$ RGS extraction region 
overlaid on the MOS\,1 image of M\,87. 

\begin{figure*}
\includegraphics[width=\textwidth]{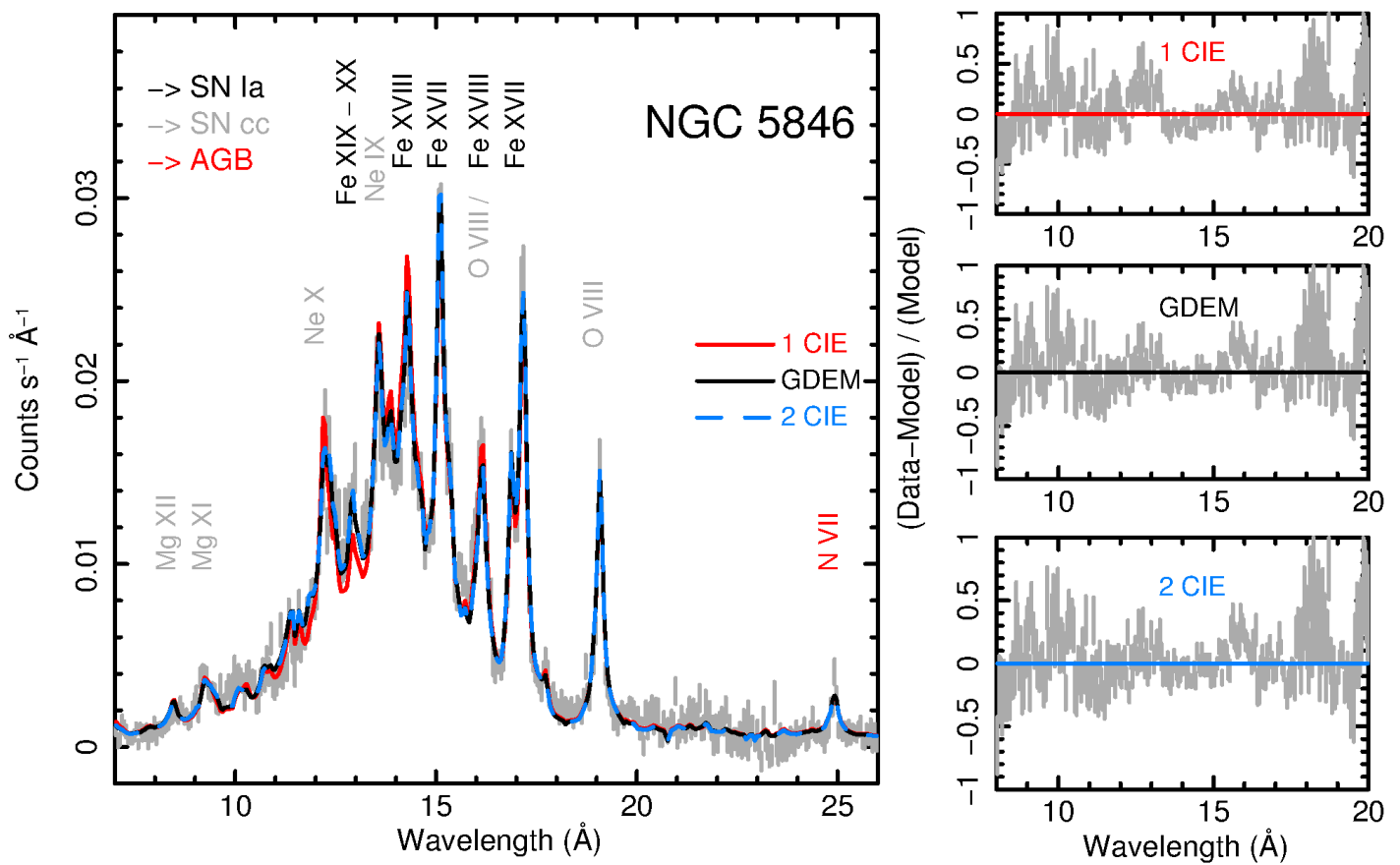}
\caption{Example stacked RGS spectrum of \object{NGC 5846}. 1T, 2T, and {\it gdem} model fits are shown.
The colored line labels indicate the most probable origin of the element, e.g., SNIa, SNcc, or AGB stars.
The residuals for the three models are shown at
the right side.}
\label{fig:ngc5846}
\end{figure*}

We also combined the RGS\,1 and 2 source spectra, the response matrices, 
and the background files extracted within the 3.4$^{\prime}$ region (see Fig.~\ref{fig:rgs_regions}) 
through the XMM-SAS task \textit{rgscombine}. These stacked spectra were only used for plotting purposes.
The spectral fits were performed simultaneously on the individual spectra. The stacked RGS spectrum of NGC 5846 is 
shown in Fig.~\ref{fig:ngc5846} as an example. We converted the spectra into the SPEX format because we used the 
SPEX\footnote{http://www.sron.nl/spex} spectral fitting package version 3.02.00 for the spectral fitting 
\citep{kaastra1996}.

Since we analyzed RGS spectra in units of counts, the errors on the data points are Poisson distributed. 
Therefore, we minimized the C-statistic \citep{cash1979} when we fit models to the RGS spectra.

\subsection{RGS spectral broadening}
\label{sec:rgs_broadening}

Since RGS is a spectrometer without a slit, the spatial extent of the source causes the measured
spectral lines to be broadened \citep[see][for a discussion about grating responses]{davis2001}. 
Photons originating from a region near the cluster center but
offset in the direction along the dispersion axis ($\Delta\Theta$ in arcmin) will be slightly 
shifted in wavelength ($\Delta\lambda$) with respect to line emission from the cluster center.
The wavelength shift is calculated using the following relation:
\begin{equation}
\Delta\lambda = \frac{0.138}{m} \Delta\Theta ~\AA, 
\end{equation}
where $m$ is the spectral order (see the XMM-Newton Users Handbook). We corrected for this effect 
by carefully constructing spatial profiles in corresponding spectral bands from MOS1 data. The MOS1 
detector coordinate DETY direction is parallel to the dispersion direction in RGS1 and RGS2, which 
allows a direct extraction of the surface brightness profile from a MOS1 image. We extracted MOS\,1 
images in detector coordinates for each exposure in the 0.5--1.8\,keV (7--25\,{\AA}) energy band. For 
each image, we extracted the surface brightness profile in the dispersion direction through the 
\textit{Rgsvprof} task, which is part of the SPEX spectral fitting package. From a MOS1 detector image, 
this task derives the cumulative spatial profile along the dispersion direction for a certain width in 
the cross-dispersion and the dispersion direction. For the width in cross-dispersion, we chose widths 
of 3.4$^{\prime}$ and 0.8$^{\prime}$. The width in the dispersion direction is set to 10$^{\prime}$, 
since the bulk of the cluster emission is contained within this radius.  

The spatial profile from \textit{Rgsvprof} was convolved with the model spectrum during spectral fitting
using the {\it lpro} model component in SPEX. The main point of this procedure is to include the 
line broadening that is due to the spatial extent of the source in the spectral model. It allows us to fit the 
broadening of the spectral lines and propagate the uncertainty in the spatial broadening into the 
uncertainties of the other fit parameters, such as the O and Fe abundances. \citet{pinto2015} showed some 
examples of spatial profiles in their Figure 2. 

\subsection{Spectral modeling}
\label{sec:specmodeling}

In order to model the multi-temperature structure in clusters \citep[see, e.g.,][]{frank2013}, we used
and compared several models available in the SPEX package.
In addition to the simple one-temperature (1CIE) and two-temperature (2CIE) models, we also used differential
emission-measure (DEM) models. In these models, emission measures 
are assumed for a range of
temperatures on a grid that follow a model or empirical parametrization of the emission measure distribution. The
empirical parametrizations are either a truncated power-law distribution ({\it wdem}, Sect.~\ref{sec:wdem}) 
or a Gaussian distribution ({\it gdem}, Sect.~\ref{sec:gdem}). For spectral simulations, we also used the 
classical cooling-flow model. In the models, we fixed the redshift to the most accurate value from optical 
observations, and we used the Galactic column densities estimated using EPIC \citep{mernier2016a}, unless stated
otherwise. We did not use literature values for the $N_{\mathrm{H}}$, because we found that $N_{\mathrm{H}}$ 
is a source of systematic uncertainty (see Section~\ref{sec:nh-bias}).

We note that by fitting X-ray spectra, it is difficult to distinguish between the different DEM model 
parametrizations, like {\it wdem} and {\it gdem}. \citet{kaastra2004b} showed that different temperature 
distributions that share the same emission-weighted average temperature and the same total emission measure
produce very similar X-ray spectra that are usually statistically indistinguishable from each other. These
DEM models do yield somewhat different abundances when fitted to spectra, therefore multi-temperature structure is a 
source of systematic uncertainty for abundances that needs to be addressed (see Section~\ref{sec:multi-t}).  

In all the DEM models, it is implicitly assumed that the abundances in the plasma are the same for all 
temperature components in the region where the spectrum was extracted. With the current spectral 
resolution, it is in most cases very hard or even impossible to resolve individual thermal components 
and to uniquely determine abundances for each temperature. Therefore, we need this assumption to obtain 
stable fit solutions. This means that the abundances that we measure are essentially emission-weighted
average abundances in the fitted region.

\subsubsection{1CIE and 2CIE modeling}

All spectra were initially fit using two temperature components (in collisional ionization equilibrium, 
CIE). The temperatures and emission measures of the two components were left to vary. When one of the components
was poorly constrained, a single-temperature or {\it gdem} model was chosen. The abundances of both
components were coupled to each other in order to be consistent with the DEM models, which assume that
all temperature components have the same abundance.

\subsubsection{WDEM model}
\label{sec:wdem}

One of the differential emission measure models we used is the so-called {\it wdem} 
model, where the emission measure, $Y = \int n_e n_H dV$, of a number of thermal 
components is distributed as a truncated power law. This is shown in Eq.~(\ref{eq:dy_dt})
adapted from \citet{kaastra2004}:
\begin{equation}
\frac{dY}{dT} = \left\{ \begin{array}{ll}
cT^{1/\alpha} & \hspace{1.0cm} \beta T_{\mathrm{max}} \le T < T_{\mathrm{max}} \\
0 & \hspace{1.0cm} T > T_{\mathrm{max}} \lor T < \beta T_{\mathrm{max}} .\\
\end{array} \right.
\label{eq:dy_dt}
\end{equation}
This distribution is cut off at a fraction of $T_{\mathrm{max}}$ that is $\beta T_{\mathrm{max}}$.
The value of $\beta$ was set to 0.1 in this study, which roughly corresponds to the lowest temperatures
that are typically detectable with the RGS. The model above is an empirical parametrization
of the DEM distribution found in the cores of cool-core clusters \citep{kaastra2004,sanders2010}. 
In this form, the limit $\alpha \to 0$ yields the isothermal model at $T_{\mathrm{max}}$. 

\subsubsection{GDEM model}
\label{sec:gdem}

Another DEM model that we used is a Gaussian differential emission measure distribution, {\it gdem}, 
in $\mathrm{log}~T$ \citep{deplaa2006}:
\begin{equation}
Y(x) = \frac{Y_0}{\sigma_{\mathrm{T}} \sqrt{2\pi }} \mathrm{e}^{-(x-x_0)^2 / 2\sigma^2_{\mathrm{T}}}.
\end{equation} 
In this equation, $x=\mathrm{log}~T$ and $x_0=\mathrm{log}~T_0,$ where $T_0$ is the average temperature 
of the distribution. The width of the Gaussian is $\sigma_{\mathrm{T}}$. Compared to the {\it wdem} 
model, this distribution contains more emission measure at higher temperatures. This model usually 
yields very similar C-statistic values as the {\it wdem} model when fitted to cluster spectra. It 
resembles the overall shape of the isobaric cooling-flow model, but without the strong emission measure
around 0.4 keV, which was not detected with XMM-Newton \citep[see, e.g.,][]{peterson2001,tamura2001a}.

\subsubsection{Cooling-flow model}
\label{sec:coolflow}

The isobaric cooling-flow model \citep[see, e.g.,][]{fabian1994} is the only physical DEM model
we used. For this DEM model, the emission measure distribution ($dY/dT$) is described by
\begin{equation}
\frac{dY}{dT} = \frac{5 \dot{M} k}{2 \mu m_H \Lambda (T)}
,\end{equation} 
where $\dot{M}$ is the cooling rate in $\mathrm{M}_{\odot}$/yr, $k$ is the Boltzmann constant, $\mu$ is the mean 
molecular weight, and $m_H$ is the mass of a hydrogen atom. The $\Lambda (T)$ stands for the cooling
function, which is pre-calculated for an abundance of 0.5 times the solar abundance \citep{kaastra2004}.
In the model, the $dY/dT$ is calculated for a grid truncated at a low temperature $T_1$ and a high 
temperature $T_n$. The temperature grid typically contains 16 bins.

\subsubsection{Updated atomic data and radiation processes}
\label{sec:newspex}

We used thermal plasma models that were developed from the orginal MEKAL code \citep{mewe1985,mewe1986},
with a major update of the Fe-L complex lines by \citet{liedahl1995}. This original code has been included in 
XSPEC, but has not been updated since then. The MEKAL development continued as an integral part of the 
SPEX code \citep{kaastra1996}, where it has been available as the default CIE model. Over the years, it 
received some updates. Although it is not available separately from SPEX, the model is built up from an 
atomic database and a set of routines that calculate the emission processes and the resulting model spectrum.
The database and the related routines are called SPEXACT\footnote{SPEX Atomic Code and Tables}. Between 1996 and 2016,
the CIE model in SPEX was updated regularly and was the default SPEX CIE model. We refer to this model 
as SPEXACT version 2.06. 

With the release of SPEX version 3.0 early in 2016, a newly developed spectral emission code became publicly 
available in the SPEX package. This code contains newly calculated atomic data and more accurate approximations 
of the emission processes in hot plasmas. For example, the radiative recombination (RR) component of the line 
emissivity was approximated by a power law in SPEXACT v1 and v2, while the true relation is slightly curved, 
which causes the oxygen abundance to be biased in certain temperature ranges \citep[see][]{mernier2016a}. 
In SPEXACT v3.02, the RR rates are updated and now produce a much more accurate oxygen abundance values.
In this paper, we mainly used SPEXACT version 3.02 to fit the spectra. The iron lines, however, were still 
calculated using SPEXACT 2.06 because the \ion{Fe}{xvii} lines are very uncertain in the models 
\citep{deplaa2012}. For this paper, we used the calculation by \citet{doron2002}, which appears to 
describe the observed \ion{Fe}{xvii} line ratios reasonably well.

\section{Results}

We show the final choice of models that were fit to the spectra in Table~\ref{tab:thermal}.
For M87 and Perseus, it was necessary to include a power-law component to 
account for emission from a central active galactic nucleus (AGN). 
Most objects clearly
needed (at least) two temperatures because we observed lines from \ion{Fe}{xvii} and \ion{Fe}{xx}. In these
cases, a two-temperature fit provides the lowest C-statistics value. For some, mainly cool, objects
like M89, we do not have enough statistics to probe the multi-temperature structure, therefore we chose a single-temperature (1CIE) model. In Abell 85, the two-temperature model provides a slight improvement to a 
single-temperature model, while the {\it gdem} model does not. In the fits, we used the best-fit N$_{\mathrm{H}}$
from the EPIC analysis \citep{mernier2016a}, except for Perseus, which benefits from a free N$_{\mathrm{H}}$ value
in the fit (see Section~\ref{sec:nh-bias}).

\begin{table}
\caption{Best-fit (multi-)temperature model for each cluster.}
\label{tab:thermal}
\begin{tabular}{ll|ll}  
{\bf Source} & {\bf Model} & {\bf Source} & {\bf Model} \\
\hline\hline                     
2A0335   & 2CIE         & HCG62    & 2CIE \\                
A85      & 2CIE         & HYDRA    & 2CIE \\                
A133     & 2CIE         & M49      & 1CIE \\                
A189     & 1CIE         & M86      & 2CIE \\                                 
A262     & 2CIE         & M87      & 2CIE+PL \\             
A496     & 2CIE         & M89      & 1CIE \\                
A1795    & 2CIE         & MKW3s    & 2CIE \\              
A1991    & 2CIE         & MKW4     & 1CIE \\                
A2029    & 2CIE         & NGC507   & 2CIE \\                
A2052    & 2CIE         & NGC1316  & 2CIE \\    
A2199    & 2CIE         & NGC1404  & 2CIE  \\               
A2597    & 2CIE         & NGC1550  & 2CIE  \\               
A2626    & 1CIE         & NGC3411  & 1CIE  \\               
A3112    & 2CIE         & NGC4261  & 1CIE  \\               
A3526    & 2CIE         & NGC4325  & 2CIE  \\               
A3581    & 2CIE         & NGC4374  & 2CIE  \\       
A4038    & 2CIE         & NGC4636  & 2CIE  \\               
A4059    & 2CIE         & NGC4649  & 1CIE  \\               
AS1101   & 2CIE         & NGC5044  & GDEM   \\              
AWM7     & 2CIE         & NGC5813  & 1CIE   \\              
EXO0422  & 2CIE         & NGC5846  & 2CIE          \\               
Fornax   & 2CIE         & Perseus  & 2CIE+NH+PL   \\                
\hline
\end{tabular}
\tablefoot{\\ CIE: Single-temperature collisional ionization equilibrium model.\\
GDEM: Gaussian differential emission measure model.\\
PL: Power-law model.\\
NH: N$_{\mathrm{H}}$ left free in fitting.
}
\end{table}

\begin{table}[t]
\caption{Oxygen and iron abundances measured with the RGS. Abundances are given with respect 
to the proto-solar abundances by \citet{lodders2009}. The errors are the statistical errors.}
\setlength{\tabcolsep}{4pt}
\begin{tabular}{lddd}
{\bf Name}      & \multicolumn{1}{c}{\bf O}     & \multicolumn{1}{c}{\bf Fe}     & \multicolumn{1}{c}{\bf O/Fe} \\
\hline\hline
2A0335     & 0.59 ; 0.05 & 0.77 ; 0.05  & 0.77 ; 0.08 \\
A133       & 0.66 ; 0.08 & 0.89 ; 0.08  & 0.74 ; 0.11 \\
A1795      & 0.35 ; 0.09 & 0.41 ; 0.06  & 0.9  ; 0.3  \\
A189       & 0.8 ; 0.3   & 0.81 ; 0.18  & 1.0  ; 0.4  \\
A1991      & 0.65 ; 0.13 & 0.78 ; 0.12  & 0.8  ; 0.2  \\
A2029      & 0.41 ; 0.06 & 0.26 ; 0.03  & 1.6  ; 0.3  \\
A2052      & 0.52 ; 0.05 & 0.63 ; 0.05  & 0.84 ; 0.10 \\
A2199      & 0.62 ; 0.16 & 0.62 ; 0.12  & 1.0  ; 0.3  \\
A2597      & 0.54 ; 0.07 & 0.47 ; 0.04  & 1.13 ; 0.18 \\
AS2626     & 1.0  ; 0.7  & 1.3  ; 0.7   & 0.8  ; 0.7  \\
A262       & 0.56 ; 0.06 & 0.72 ; 0.06  & 0.78 ; 0.10 \\
A3112      & 0.51 ; 0.06 & 0.59 ; 0.05  & 0.87 ; 0.12 \\
A3526      & 0.82 ; 0.04 & 1.22 ; 0.05  & 0.67 ; 0.05 \\
A3581      & 0.47 ; 0.04 & 0.54 ; 0.03  & 0.86 ; 0.09 \\
A4038      & 0.66 ; 0.14 & 0.61 ; 0.11  & 1.1  ; 0.3  \\
A4059      & 0.58 ; 0.09 & 0.86 ; 0.10  & 0.68 ; 0.13 \\
A496       & 0.60 ; 0.06 & 0.67 ; 0.05  & 0.89 ; 0.12 \\
A85        & 0.55 ; 0.07 & 0.70 ; 0.07  & 0.77 ; 0.12 \\
AS1101     & 0.32 ; 0.04 & 0.42 ; 0.03  & 0.76 ; 0.11 \\
AWM7       & 0.59 ; 0.08 & 0.49 ; 0.05  & 1.20 ; 0.19 \\
EXO0422    & 0.65 ; 0.15 & 0.70 ; 0.13  & 0.9  ; 0.3  \\
Fornax     & 0.54 ; 0.06 & 0.80 ; 0.07  & 0.68 ; 0.10 \\
HCG62      & 0.45 ; 0.05 & 0.56 ; 0.04  & 0.80 ; 0.11 \\
HYDRA      & 0.35 ; 0.05 & 0.32 ; 0.04  & 1.1  ; 0.2  \\
M49        & 0.61 ; 0.06 & 0.62 ; 0.04  & 0.99 ; 0.12 \\
M86        & 0.51 ; 0.08 & 0.40 ; 0.04  & 1.27 ; 0.25 \\
M87        & 0.62 ; 0.14 & 0.60 ; 0.13  & 1.0  ; 0.3 \\
M89        & 0.49 ; 0.12 & 0.24 ; 0.04  & 2.0  ; 0.6  \\
MKW3s      & 0.37 ; 0.10 & 0.52 ; 0.06  & 0.7  ; 0.2  \\
MKW4       & 0.86 ; 0.13 & 1.07 ; 0.07  & 0.80 ; 0.13 \\
NGC1316\tablefootmark{a} &      &       & 1.9  ; 0.3  \\
NGC1404    & 0.65 ; 0.13 & 0.56 ; 0.07  & 1.2  ; 0.3  \\
NGC1550    & 0.59 ; 0.07 & 0.80 ; 0.07  & 0.73 ; 0.11 \\
NGC3411    & 0.9  ; 0.3  & 1.3  ; 0.2   & 0.7  ; 0.2  \\
NGC4261    & 0.48 ; 0.08 & 0.37 ; 0.04  & 1.3  ; 0.3  \\
NGC4325    & 0.44 ; 0.11 & 0.63 ; 0.08  & 0.70 ; 0.19 \\
NGC4374    & 0.63 ; 0.11 & 0.43 ; 0.07  & 1.5  ; 0.4  \\
NGC4636    & 0.63 ; 0.05 & 0.59 ; 0.03  & 1.07 ; 0.10 \\
NGC4649    & 0.63 ; 0.05 & 0.66 ; 0.03  & 0.96 ; 0.10 \\
NGC5044    & 0.56 ; 0.03 & 0.54 ; 0.02  & 1.02 ; 0.07 \\
NGC507     & 1.1  ; 0.3  & 1.4  ; 0.2   & 0.8  ; 0.3  \\
NGC5813    & 0.63 ; 0.05 & 0.59 ; 0.03  & 1.07 ; 0.10 \\
NGC5846    & 0.81 ; 0.07 & 0.66 ; 0.04  & 1.22 ; 0.14 \\
Perseus    & 1.19 ; 0.18 & 0.88 ; 0.14  & 1.4  ; 0.3  \\
\hline
$\mu$\tablefootmark{b}  & 0.551 ; 0.010 & 0.556 ; 0.007 & 0.853 ; 0.018  \\
$\sqrt{\mathrm{Var}}$ & \multicolumn{1}{c}{0.22} &  \multicolumn{1}{c}{0.52} & \multicolumn{1}{c}{0.34} \\
$\chi^2$ & \multicolumn{1}{c}{175/43} & \multicolumn{1}{c}{702/43} & \multicolumn{1}{c}{102/43} \\
\hline
\end{tabular}
\tablefoot{
\tablefoottext{a}{Reference atom in CIE model set to iron.}
\tablefoottext{b}{Weighted mean abundance.}
}
\label{tab:abundances}
\end{table}

The final fit results for oxygen and iron obtained from the RGS are listed in Table~\ref{tab:abundances}. Since
absolute abundances can be more sensitive to systematic effects than relative abundances, we also calculated 
the O/Fe ratio for comparison. The weighted mean abundances for O and Fe are 0.551$\pm$0.010 and 
0.556$\pm$0.007, respectively. The O and Fe values show considerable scatter. A calculation of the 
variance yields a value of 0.22 for O and 0.52 for Fe. The scatter in the ratio O/Fe is 0.34. 

We can assume that the statistical errors on the measured O and Fe abundances are approximately 
normally distributed, which means that we can use $\chi^2$ statistics when we fit a model to these abundances.
When we assume that the parent population of clusters has a constant O/Fe ratio, a fit with a constant value to the
abundances yields a $\chi^2$ of 102 / 43 d.o.f., which is formally not acceptable, but much smaller than 
the $\chi^2$ of the individual O and Fe abundances.

\begin{figure}[!t]
\includegraphics[width=\columnwidth]{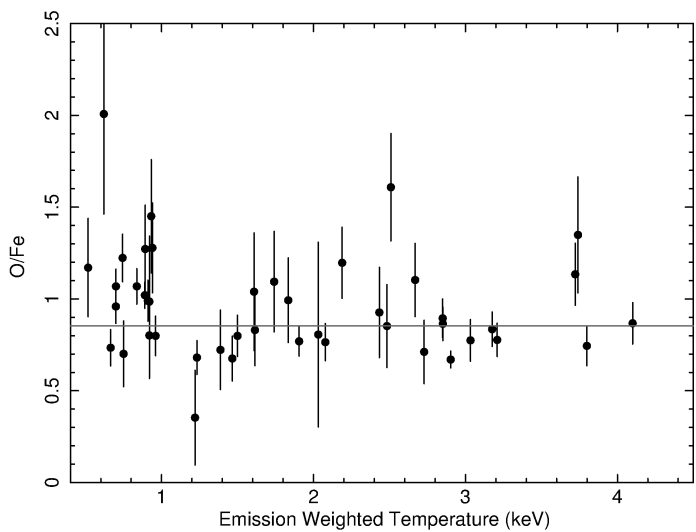}
\caption{O/Fe ratio plotted against the emission weighted temperature determined from RGS. The dark grey 
line shows the error weighted mean of the sample.}
\label{fig:kt-ofe}
\end{figure}

\begin{figure}[!t]
\includegraphics[width=\columnwidth]{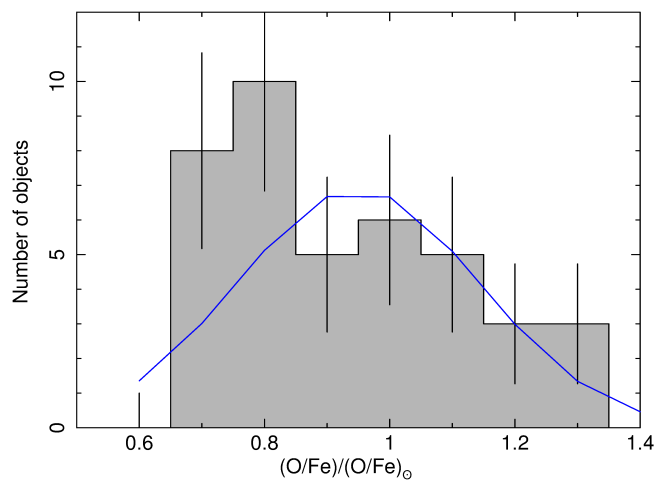}
\caption{Histogram of the measured O/Fe ratios in the CHEERS sample. The blue line shows a Gaussian fit to 
the distribution.}
\label{fig:hist}
\end{figure}

The weighted average of the measured O/Fe ratio is 0.853 $\pm$ 0.018, which is not the same as the ratio between 
the mean absolute oxygen and iron abundance. These are just two different estimators, and it is not expected 
that they yield the same result. Since we used linear abundance ratios, it cannot be expected either
that the average $<$O/Fe$>$ is the same as $<$Fe/O$>^{-1}$. When we plot the
O/Fe ratio as a function of the dominant plasma temperature (see Fig.~\ref{fig:kt-ofe}), the points
with the smallest error bars appear to cluster around a O/Fe value of 0.8. The points above 1.0 have
larger error bars, which is partly due to the error propagation. This means that objects with a lower O/Fe 
are assigned a slightly higher weight than clusters with a high O/Fe. In a histogram
of the O/Fe values (see Fig.~\ref{fig:hist}),  a hint of a tail toward higher abundance 
ratios is visible. When we fit a single Gaussian to the histogram, we find the center of the Gaussian at
0.95 $\pm$ 0.04 and a width of 0.19 $\pm$ 0.03. The $\chi^2 / d.o.f.$ of the distribution is 9 / 6, which 
is formally acceptable. The number of objects in our sample is too low to allow us to statistically 
distinguish between more complicated shapes of the O/Fe distribution, for example, a bimodal distribution.

\section{Fitting biases}
\label{sec:fitbias}

The accuracy of the measured abundances need to be studied carefully because several systematic
effects can influence the value measured in the spectral fit. First, we study the effect of spatial 
line broadening on the abundance measurement that is due to the slitless nature of the RGS (see Section~\ref{sec:bias}). 
Since the line profile shows the spatial distribution of the emissivity of the ion, each line has in principle a 
different broadening, which is not corrected for in the spectral fit. Second, the choice of the
multi-temperature model fixes the shape of the emission measure distribution, which may not reflect the true 
distribution in the gas and bias the abundance measurement (see Section~\ref{sec:multi-t}). In Section~\ref{sec:nh-bias}
we study the influence of the assumed $N_{\mathrm{H}}$ value on the abundance measurement. Finally, thermal 
plasma codes also contain uncertainties that affect the abundance measurement (see Section~\ref{sec:sysnewatom}). 
In Sections~\ref{sec:biassim} and \ref{sec:biastot} we attempt to estimate the effect of the systematic biases
on the abundance result.

For each bias effect, we simulate spectra using the RGS response matrices. Poisson noise is not added
to the simulated spectra, which means that the value of each data point is the exact mean number of counts 
expected for that bin in a 100 ks observation. The error on each data point is set to be the square root of 
the expected number of counts. We checked that the results of
the fit are the same as long as the simulated spectra have a minimum 
quality level ($\gtrsim$10 000 counts). We do not need
many Monte Carlo simulations in this case because we know the expected value for each simulated spectral 
bin exactly: it is the model value for that bin. We also know the expected variance for each bin, which is the 
square root of the model value for a given exposure time. If Poisson noise were added, we would need to average out 
the statistical fluctuations through many Monte Carlo runs to obtain an approximation of the input value, which 
we already know exactly. Since we wish to compare models with each other, statistical fluctuations are not relevant. 
The average difference between the best-fit model values and the consequences for the fitted parameters are important,
but do not depend on the statistical noise. Only the statistical weight of a data point in the fit minimization matters 
for the comparison, which is included by the error bars on the simulated data points. We chose this method over a Monte Carlo method with Poisson noise because 
it saves much unnecessary computation time.      

When we compared the SPEX CIE model to APEC version 3.0.1, we 
used XSPEC to simulate spectra using the APEC model. Since we need to use the same
solar abundance table in both SPEX and XSPEC for the comparison and it does not really matter which one, 
we chose the \citet{lodders2003} solar photospheric abundances that are available in both packages. 
In the simulations, solar abundances are assumed
for both O and Fe (always 
with respect to the same solar abundance table as used in the corresponding fits). In the cooling-flow 
model simulation, the low-temperature cutoff is set at 0.5 keV, and in the {\it wdem} models, the cutoff 
is set at 0.2 keV and $\alpha$ to 2.0.

\subsection{Bias that is due to spatial line-broadening}
\label{sec:bias}

Owing to the nature of the RGS, spectral lines of extended sources are broadened with a width 
that depends on the spatial emissivity distribution of the respective ion on the sky 
(see Sect.~\ref{sec:rgs_broadening} for an explanation of spatial line broadening). This means that 
essentially two factors determine the width of a line in the
RGS: the radial abundance distribution, which
sets the amount of elements in the gas as a function of radius, and the radial temperature distribution, which sets
the abundance of each ion following the ionization balance at that temperature. These two parameters 
govern the line emissivity as a function of radius. Temperature and abundance gradients in the cores 
of clusters therefore can cause lines of different ions to show different line broadening profiles 
in RGS spectra. Because abundance profiles of different elements are usually not very 
different from each other \citep[e.g.,][]{mernier2017}, the main cause for differences in line broadening is a
steep temperature profile that causes the ionic fractions of ions to vary strongly with radius. This was confirmed
by \citet{pinto2016}, who found smaller line widths in cool-temperature components  
and broader line widths in hotter temperature components in the cores of elliptical galaxies. 
 
In some clusters, the variation in broadening between lines of different ions can be a significant 
effect. In extreme cases, the \ion{O}{VIII} line width can be about a factor of 3 larger than the average 
widths of the iron lines \citep{deplaa2006}. Since current spectral fitting 
programs cannot fully model the widths of each line, the spectra are fit with an average width. The question 
is whether this introduces a systematic bias in the abundance determination.   

\begin{figure}
\includegraphics[width=1.0\columnwidth]{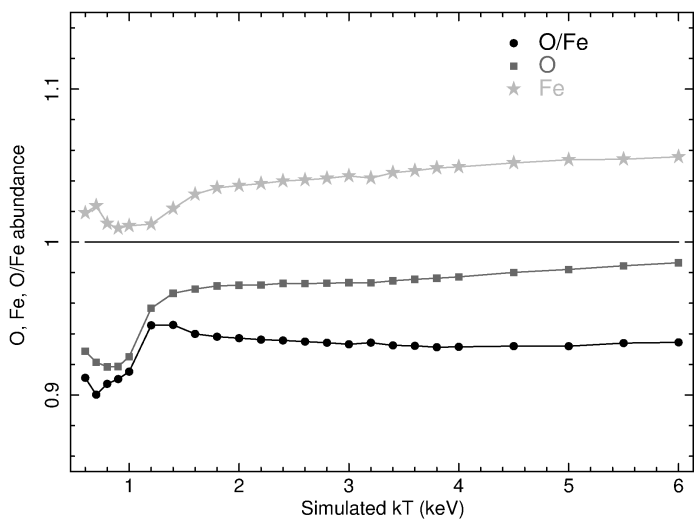}
\caption{O, Fe, and O/Fe abundance results for fits to simulated spectra of a range of temperatures. 
In the simulated spectra, we set the width of the oxygen lines to be twice the width of the Fe lines.
In the fit, this difference in width is not fit and is assumed to be the same for all lines.
The squares and stars show the absolute O and Fe abundance, respectively. The circles show the O/Fe ratio. All 
measurements are compared to the input value of 1.}
\label{fig:wdem-width}
\end{figure}

To estimate the effect of line broadening on the measured O/Fe abundance, we simulated RGS spectra for a range 
of temperatures from 0.6 keV to 6.0 keV. The simulated spectra consisted of the addition of two model spectra
with the same temperature and normalization, but with a different broadening profile. For the first model 
spectrum, we set the O, Ne and Mg abundance to twice solar and the Fe abundance to zero. We broadened this 
spectrum with a typical spatial profile for a cluster. In the second model spectrum, the O, Ne and Mg abundance 
were set to zero and Fe was set to twice solar. This model spectrum was broadened with a profile with the 
same overall shape, but scaled to half the width of the first spectrum. When the two spectra were added, the total 
spectrum mimicked a spectrum with twice the normalization of the individual components and with the average 
abundance of both components (i.e., once solar). We chose an average abundance of once solar in the 
simulations to facilitate recognizing the relative difference between the input and output abundance. 

We divided the detectable elements in the RGS into two groups (we assigned O, Ne, and Mg with a broadening   
twice larger than the broadening of the Fe lines) because the broadening is mainly determined by the strongest lines of Fe and O. The Ne and
Mg lines are not that strong and have a lower weight in the determination of the broadening. Given the similar
origin of O, Ne, and Mg (core-collapse supernovae) and their comparable atomic weight, we assumed that the line widths
of O, Ne, and Mg are very similar to each other and coupled their widths in this simulation to the width of oxygen.
The factor of two difference in line width is based on the typical ratio between the line widths of the observed 
O and Fe lines in different clusters, which in practice varies from 1 (not significantly different) up to a factor of 
3 in an extreme case \citep{deplaa2006}.

We fit the simulated spectrum for each temperature bin with a single-temperature model, but now 
with a single line-broadening component. The resulting O, Fe, and O/Fe abundance measurements are shown in 
Fig.~\ref{fig:wdem-width}. We find a bias of about $-10\%$ in the O/Fe abundance ratio, which weakly depends 
on the plasma temperature. Because the fit to the simulated spectra was now constrained to a single broadening profile,
it tried to find a weighted-average width between the O and Fe line widths. For O, the model line width was
smaller than the simulated width, which means that some of the line flux in the wings of the line was effectively attributed 
to the continuum. A lower line/continuum ratio leads to an underestimation of the O abundance. For Fe, this works in
the opposite way. Because the line model is broader than the simulated line, continuum photons are attributed to 
the wing of the line, which leads to a higher model line flux and hence to an overestimation of the Fe abundance. 
Around 1 keV, the Fe line flux is increased through the higher contribution of \ion{Fe}{xvii} lines. Therefore, Fe
has a relatively large weight in the determination of the line width, and the measured Fe abundance is therefore
closer to the input value (at the cost of a larger bias in O). For higher temperatures, the balance between the Fe and 
O line fluxes changes in favor of O, which means that the bias in O decreases while the bias in Fe increases.   

For verification purposes, we also simulated and fit spectra without line broadening. The differences between the input and output values were then smaller than 1\%. Therefore, we can fully
attribute 
the biases observed in Fig.~\ref{fig:wdem-width} to line broadening effects. 
In reality, the effect of line broadening may be slightly different. In the simulation, the lines are 
broadened by a convolution, while in RGS observations, the core of the line is as strong as it should be and 
the wings of the lines are enhanced by the line emission from the regions around the core. Our simulations 
show the typical magnitude of the systematic differences in abundances that are to be expected, but in a real case, the 
systematic difference may be in the other direction. 

\subsection{Bias that is due to multi-temperature structure}
\label{sec:multi-t}

It is clear that because of the relatively large field of view of the RGS and the thermally complicated nature of
cooling cores, the measured spectra might likely not consist of a single-temperature spectrum. Their true
multi-temperature structure is, however, not uniquely constrained. The only physical multi-temperature model that we 
have is the cooling-flow model (see Section~\ref{sec:coolflow}), but it was found to describe the 
first XMM-Newton spectra of clusters not very well \citep[e.g.,][]{peterson2001,tamura2001a}. Therefore, we used 
other (empirical) models described in Section~\ref{sec:specmodeling} in an attempt to approximate the 
emission-measure distribution with a power law ({\it wdem}) or a Gaussian function ({\it gdem}). Since the 
true distribution is not known and the emissivity of lines depends on the temperature, imperfections in 
the multi-temperature approximation may bias the measured abundances. We therefore tried to determine 
biases in abundances for different combinations of multi-temperature models to estimate the typical
magnitude of the bias that is due to multi-temperature effects. 

\begin{figure}
\includegraphics[width=1.0\columnwidth]{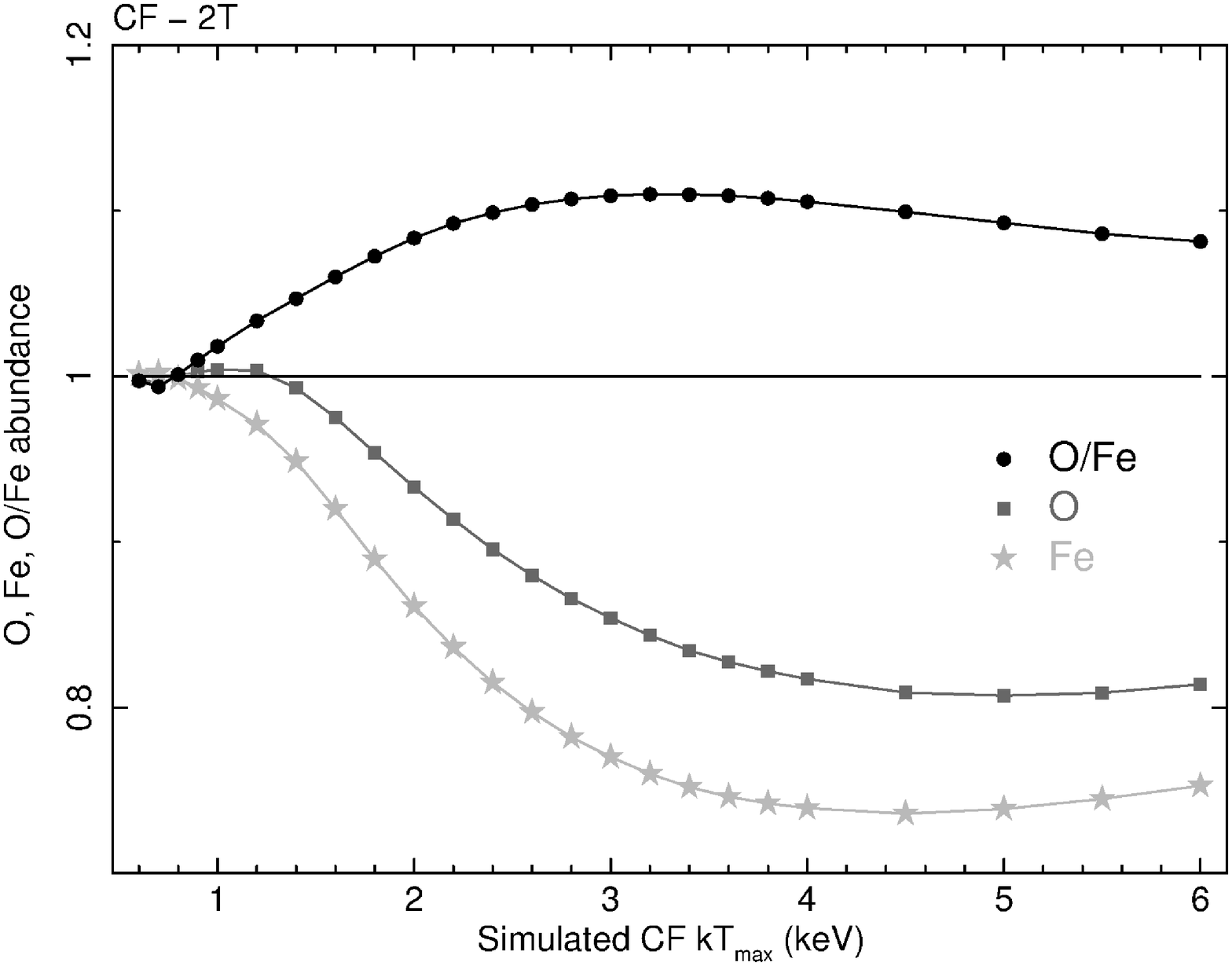}
\caption{Results from two-temperature fits to simulated RGS cooling-flow spectra for a range of (maximum) 
temperatures. The measured O, Fe, and O/Fe abundances are shown and compared to their input value of once solar.}
\label{fig:cf-2t}
\end{figure}

In Fig.~\ref{fig:cf-2t} we show how O and Fe abundance measurements are biased when we simulate an input 
spectrum using a cooling-flow model and fit it with two single-temperature models. The input cooling-flow
model had a low $kT$ limit of 0.5 keV because otherwise we would
have created strong \ion{O}{vii} and \ion{Fe}{xvii} 
lines that are not observed at this high emissivity \citep{peterson2001}. At low temperatures, the
two-temperature model reproduced the O and Fe abundances well. For higher temperatures, however, the O/Fe was 
overestimated because the Fe abundance was underestimated.   

\begin{figure}
\includegraphics[width=1.0\columnwidth]{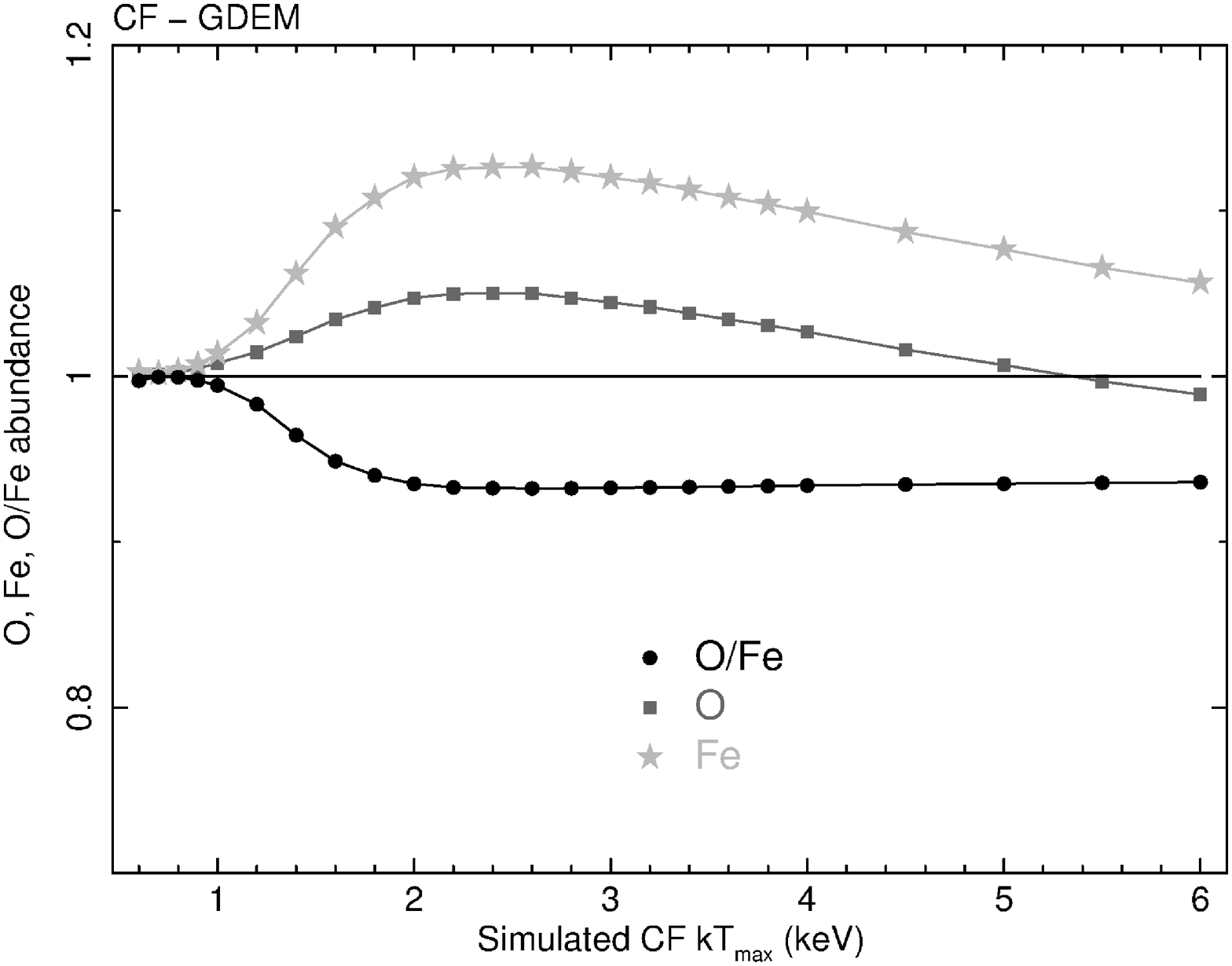}
\caption{Results from {\it gdem} fits to simulated RGS cooling-flow spectra for a range of (maximum) 
temperatures. The measured O, Fe, and O/Fe abundances are shown and compared to their input value of once solar.}
\label{fig:cf-gdem}
\end{figure}

When we performed the same experiment, but fit the simulated spectra from the cooling-flow model with a {\it gdem}
model, we detected a bias in the other direction. Again, the
input cooling-flow model had a low $kT$ limit of 0.5 keV. 
Figure~\ref{fig:cf-gdem} shows the results. The {\it gdem} model fits the O/Fe abundance well below 1 keV, but 
above this temperature, the measured abundances deviate from each other. The resulting O/Fe abundance is 
underestimated by about 10\% in this case.

\begin{figure}
\includegraphics[width=1.0\columnwidth]{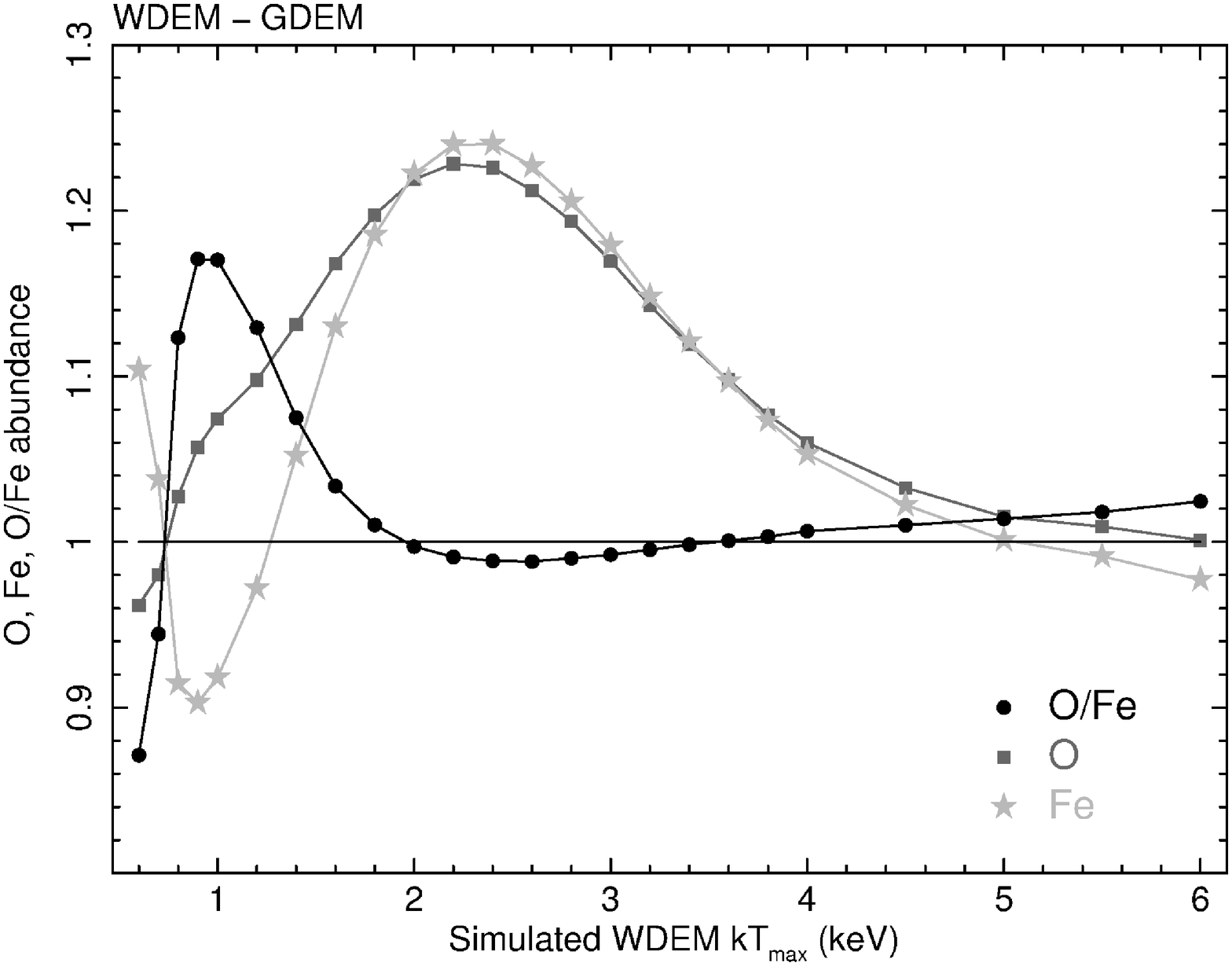}
\caption{Results from {\it gdem} fits to simulated RGS {\it wdem} spectra for a range of (maximum) 
temperatures. The measured O, Fe, and O/Fe abundances are shown and compared to their input value of once solar.}
\label{fig:wdem-gdem}
\end{figure}

In a similar way, we also compared the {\it wdem} and {\it gdem} models. The input DEM parameters for the 
{\it wdem} simulated spectra were $\alpha=2$ and $\beta=0.2$, which are typical observed values for clusters. 
Figure~\ref{fig:wdem-gdem} shows results from the simulated {\it wdem} spectra fit with {\it gdem} models. In this 
case, the variation in the results is much larger with temperature. The most significant variations are seen at the 
low-temperature end, where O and Fe are biased in opposite directions. Around 1 keV, the bias is about $20\%$ in 
the O/Fe ratio. However, above 2 keV, the bias in the O/Fe drops to a few percent.

For the line-broadening bias (see Section~\ref{sec:bias}), it is relatively easy to qualitatively 
explain the observed biases. For the multi-temperature biases that we have estimated in this section, however, it is 
far more difficult. The reason is that normalization, temperature, and abundance can partly compensate for 
each other, especially for temperature components that are not dominant. When \ion{Fe}{xvii} lines are detected, for 
example, which indicates the presence of gas with a temperature around 0.5$-$0.7 keV, the fit can either try to 
increase the normalization of the low temperatures in the DEM model, lower the central temperature of the DEM, or 
increase the Fe abundance. Changing these parameters also affects other bands in the spectrum through the continuum. 
Therefore, the fit result is the result of a complicated interplay between the assumed temperature distribution
and the abundances. 

The experiments above show that the bias in the O/Fe ratio and the individual abundances are diverse. 
We only performed a small subset of tests, which provides a general idea of the accuracy, but not a precise 
measure. It is therefore difficult to know the bias that is due to multi-temperature structure exactly. 
Based on this experiment, we can only estimate that the accuracy of the O/Fe abundance ratio is about $10-20\%$.

Since we do not observe a strong relation between the O/Fe ratio and temperature in the CHEERS sample, it 
does not appear that we systematically used an inappropriate DEM distribution to fit the spectra. However, we do observe
a significant scatter in the measured abundances values. We speculate that the multi-temperature structure 
varies between objects and that we are unable to model this structure precisely, which leads to random 
biases in the abundances. Random biases of about $10-20\%$, as we typically find in our simulations, could 
be one of the main sources of the scatter that we observe in the O and Fe abundances.

\subsection{Bias that is due to uncertainties in the broad-band continuum}
\label{sec:nh-bias}

Since line strengths are measured with respect to the continuum, uncertainties in the broad-band continuum
can significantly affect the measured abundances. The continuum consists of multiple continua from the CIE model,
background components, possible AGN power-law emission, and is absorbed by the Galactic interstellar medium (ISM).
For most of the objects in the CHEERS sample, AGN power-law emission and background components do not play a major 
role because we focus on the bright inner cluster cores where the thermal emission in most cases dominates the AGN
emission and the background. In this section, we concentrate on the uncertainties that are due to the Galactic $N_\mathrm{H}$ and
discuss the biases that are due to the CIE model in Sect.~\ref{sec:sysnewatom}.

The absorption column that is due to the ISM is usually quantified using the column density of atomic hydrogen 
($N_\mathrm{H}$), which is determined using radio surveys \citep[e.g.,][]{kalberla2005}. However, in high-density 
regions, molecular hydrogen and dust can also significantly contribute to the absorption \citep{willingale2013}.
The X-ray determined $N_\mathrm{H}$ in clusters of galaxies is not always consistent with the estimates derived 
from radio and other wavelength bands. This is also partly due to calibration uncertainties in the X-ray instruments 
and uncertainties in the solar abundance table \citep{schellenberger2015}. In the EPIC analysis of the CHEERS
sample \citep{mernier2016a}, the X-ray measured $N_\mathrm{H}$ sometimes deviates from the $N_\mathrm{H}$ 
estimated by the tool provided by \citet{willingale2013}, which affects the abundances.

\begin{figure}
\includegraphics[width=1.0\columnwidth]{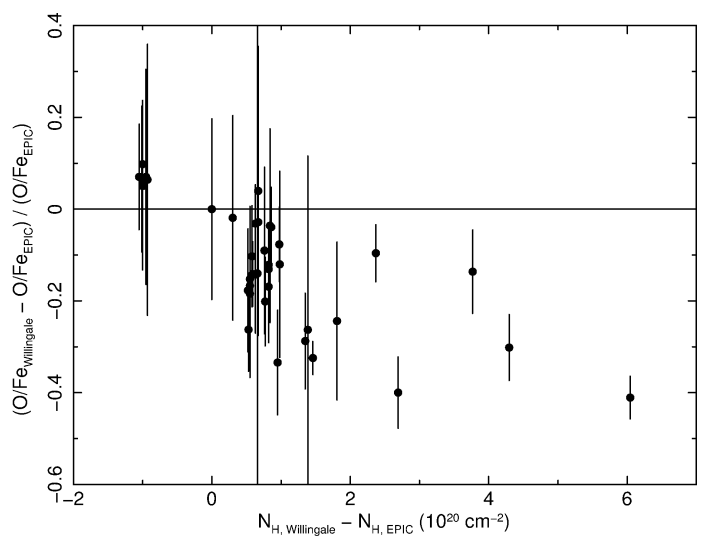}
\caption{Relative change in O/Fe as a function of difference in $N_\mathrm{H}$. Fits with $N_\mathrm{H}$ values 
of \citet{willingale2013} are compared to best-fit $N_\mathrm{H}$ values using EPIC \citep{mernier2016a}.}
\label{fig:ofe-nh}
\end{figure}

To check the effect of varying $N_\mathrm{H}$ on the RGS measured abundances, we performed the spectral fitting
twice. The first fit assumed the $N_\mathrm{H}$ using the $N_\mathrm{H}$ tool by \citet{willingale2013}, and
in the second fit, the best-fit $N_\mathrm{H}$ from EPIC \citep{mernier2016a} was assumed. For the Perseus cluster,
the $N_\mathrm{H}$ was left free because the EPIC-determined $N_\mathrm{H}$ does not fit the RGS data well. 
In Fig.~\ref{fig:ofe-nh} we show the relative difference between the O/Fe ratio measured using the \citet{willingale2013}
$N_\mathrm{H}$ values and the ratio measured using the $N_\mathrm{H}$ determined with EPIC. The plot shows a 
negative trend with increasing absolute $N_\mathrm{H}$. The bias in the O/Fe abundance ratio could increase to 
$\sim$40\% at maximum in rare cases where the absorption column is high ($\gtrsim$ 10$^{21}$ cm$^{-2}$), like for Perseus. 
The reason that we see a decreasing trend with increasing differences between the \citet{willingale2013} and
EPIC determined $N_\mathrm{H}$ values is that a higher $N_\mathrm{H}$ forces the fit to increase the continuum
of the spectral model to fit the data, which affects the line/continuum ratio such that it becomes lower. 
Since the strongest O line is located at a lower energy than the Fe-L complex, it is more strongly affected by 
$N_\mathrm{H}$ variations, and thus we observe a decreasing trend.

\begin{table}
\caption{Results of two {\it gdem} fits to the EPIC pn spectrum of the Perseus cluster using different effective areas. 
The PN calibration column shows the result for the original effective area, and the ACIS calibration column shows
the results using a modified effective area that assumes that
the Chandra ACIS calibration is correct.}
\begin{tabular}{lcc}
{\bf Parameter} & {\bf PN calibration} & {\bf ACIS calibration} \\
\hline\hline
$N_\mathrm{H}$ (10$^{21}$ cm$^{-2}$) & 1.349$\pm$0.003 & 1.398$\pm$0.003 \\
$kT$ (keV)      & 5.028$\pm$0.017    & 5.562$\pm$0.012 \\
$\sigma_{T}$    & 0.368$\pm$0.002    & 0.292$\pm$0.004 \\
O/Fe            & 0.76$\pm$0.03      & 0.82$\pm$0.03 \\
\hline
\end{tabular}
\label{tab:perseus}
\end{table}

The combination of uncertainty in $N_\mathrm{H}$ and calibration uncertainty in the soft X-ray band can set a 
significant bias on the abundance determination. The effective area calibration of the RGS and EPIC is estimated 
to be accurate within $\sim 5\%$\footnote{See XMM-SOC-CAL-TN-0018 and XMM-SOC-CAL-TN-0030 at 
https://www.cosmos.esa.int/web/xmm-newton/calibration-documentation.} 
and the systematic uncertainty in the $N_\mathrm{H}$ values of \citet{willingale2013}
are estimated to be between $\sim 8-16\%$. 
As a test of the effect of the calibration uncertainty and the $N_\mathrm{H}$ value on the O/Fe abundance,
we fit the EPIC pn spectrum of Perseus with a {\it gdem} model. The Perseus cluster shows the largest deviation in
Fig.~\ref{fig:ofe-nh} and has one of the highest temperatures of the sample. Since the effect of the effective area 
calibration increases with cluster temperature \citep{schellenberger2015}, the Perseus cluster is a conservative 
choice for this test. We fit the EPIC spectrum of Perseus twice: first, using the 
original effective area file of EPIC pn, and second, using a modified effective area that assumes that the Chandra 
ACIS calibration is correct \citep[modified using the MODARF tool,][]{schellenberger2015}. In the fits, the $N_\mathrm{H}$ was left 
free. The results are shown in Table~\ref{tab:perseus}. The change in effective area between EPIC pn and ACIS 
appears to mainly affect the measured temperature structure. The ACIS temperature is 10\% higher than 
the pn temperature. The $N_\mathrm{H}$ is hardly affected by the change in effective area, and 
its value is close to the \ion{H}{i} value from radio observations \citep[1.38 10$^{21}$ cm$^{-2}$,][]{kalberla2005}.
The effect on the O/Fe ratio is modest with a difference of $\sim$8\%. From this test, we conclude that uncertainties 
in the calibration can be compensated for by changing the model parameters, for example, the temperature in this case, which
in turn can affect the abundance determination. It is, however, difficult to draw more general conlcusions from this 
test because the compensation effects in the spectral fit can be different for each cluster or instrument.

The test shows that fixing $N_\mathrm{H}$ to the value of \citet{willingale2013} (2.12 10$^{21}$ cm$^{-2}$) would
bias the fit substantially, since both the pn and ACIS calibration favor the value of \citet{kalberla2005}.
Especially for abundance determinations, it is most important that the
continuum is estimated properly. Therefore, it is advisable to fit the $N_\mathrm{H}$, instead of fixing it to 
a literature value, which could limit the freedom of the fit to accommodate for a mismatch between the observed and 
modeled continuum. In practice, we limited the $N_\mathrm{H}$ during the fit to the range between the \ion{H}{i} value 
of \citet{kalberla2005} and the \citet{willingale2013} value, to avoid the fit to optimize to unphysical values. 
The range is slightly extended on both the low and high end with 5$\times$10$^{19}$ cm$^{-2}$ and 1$\times$10$^{20}$ 
cm$^{-2}$ , respectively, to account for the uncertainties in the $N_\mathrm{H}$ values \citep[see][for details]{mernier2016a}.

\subsection{Bias that is due to atomic database and spectral modeling accuracy}
\label{sec:sysnewatom}

The accuracy of spectral models is generally hard to assess because it requires expert knowledge of atomic physics 
and radiation processes. In X-ray astrophysics, there are two groups actively developing codes to model soft X-ray 
emission from thermal plasmas in collisional ionization equilibrium (CIE): the group developing 
APEC/ATOMDB\footnote{http://www.atomdb.org} , and the group developing SPEX. 
The bases for these codes were originally developed in the 1970s and 1980s, but in recent years, they have been 
upgraded in preparation for new instruments dedicated to high-resolution spectroscopy. Although both groups 
rely on atomic data from similar sources, the radiation processes can be complicated, and different assumptions or 
approximations can result in differences in measured abundances. 

\begin{figure}
\includegraphics[width=1.0\columnwidth]{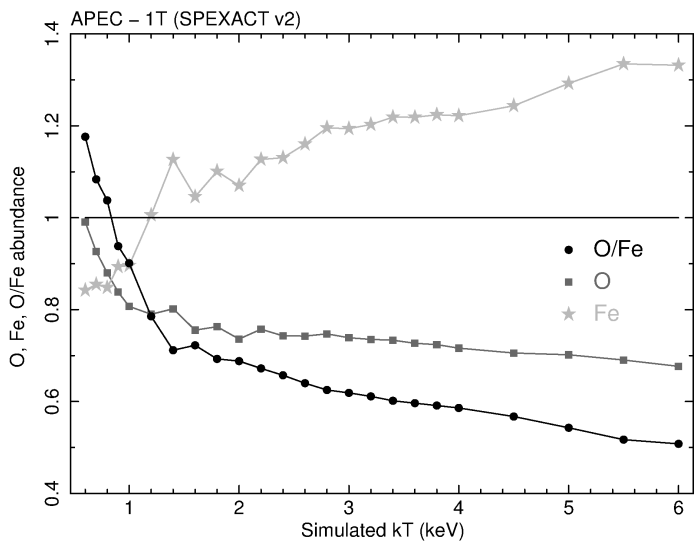}
\caption{Results from one-temperature CIE fits with SPEXACT v2 to simulated APEC spectra for a range of 
temperatures. The measured O, Fe, and O/Fe abundances are shown and compared to their input value of once solar.}
\label{fig:apec-1t}
\end{figure}

In this section, we compare the spectrum as calculated by the APEC v3.0.1 code with the default SPEXACT v2 code 
in SPEX. We also compare APEC to the new SPEXACT v3 code (see Sect.~\ref{sec:newspex}). In this comparison, we 
simulated spectra for a range of temperatures using 
APEC \citep[and][solar photospheric abundances]{lodders2003}. These spectra were subsequently fit to SPEX 
models using the same abundance set, and the normalization, temperature, and the abundances of the relevant 
elements were left free to vary. In Fig.~\ref{fig:apec-1t} we show the results for the comparison of the 
APEC spectra with SPEXACT v2 spectra. There appears to be a strong bias in the O/Fe abundance, especially 
above 1 keV, where the O/Fe ratio is about 50\% of the original value. The main origin of this difference is 
the crude approximation of the radiative recombination process in the original MEKAL model \citep{mao2016}. 
Our RGS results for oxygen are corrected for this effect.     

\begin{figure}
\includegraphics[width=1.0\columnwidth]{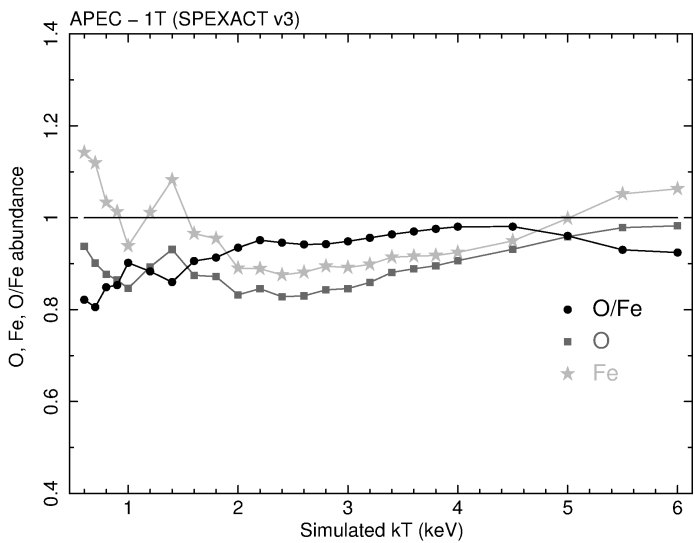}
\caption{Results from one-temperature CIE fits with SPEXACT v3 to simulated APEC spectra for a range of 
temperatures. The measured O, Fe, and O/Fe abundances are shown and compared to their input value of once solar.}
\label{fig:apec-1t-new}
\end{figure}

In the SPEXACT v3 code, the issue with the radiative recombination is fixed and we can compare the
APEC/ATOMDB v3.0.1 and SPEXACT v3.02 to see what the remaining uncertainties are in O/Fe. Figure~\ref{fig:apec-1t-new} shows the fit 
results for the same simulated APEC spectra, but now fit with the latest SPEX model. The strong trend with
temperature that was visible in the previous plot for the SPEXACT v2 code has become less pronounced. However, a 10-20\% discrepancy between APEC v3.0.1 and SPEX v3.02.00 remains. The origin of these differences
are subject of study by the APEC and SPEX teams and include biases that are due to, for example, differences in 
included atomic data and differences in modeled radiation processes. We expect that only new high-resolution 
X-ray spectrometers like SXS on board Hitomi \citep{takahashi2010} and the X-ray Astronomy Recovery Mission (XARM)
or the X-IFU instrument on board the future X-ray observatory ATHENA \citep{nandra2013} will be able to 
provide a sufficient benchmark to improve these spectral models.

Thanks to this study, we were able to find and correct a source of bias in the old thermal models of
MEKAL \citep{mao2016}. This reduced the difference in the O/Fe ratio from 50\% to at most 10$-$20\%. 
Differences between the SPEXACT v3 CIE model and APEC/ATOMDB v3.0.1 remain, which can likely be 
attributed to differences in the amount of spectral lines in the database and differences in the 
implementation of radiative processes. 
This case shows that deep observations help to identify and solve systematic biases in the results.

\subsection{Effect of biases on a simulated sample}
\label{sec:biassim}

The systematic errors on O/Fe that we estimated in the previous sections cannot simply be added  
because the bias in O/Fe has a highly nonlinear dependence on the observation and cluster parameters.
For each cluster, the total bias may turn out to be different. One possible way to estimate the total
effect of the biases is to simulate a number of clusters, add as many biases as possible, and measure
the effect on O/Fe. Therefore we simulated a set of 50 RGS spectra of 100 ks using SPEXACT v3. 
In the simulated spectra, the input abundances for O and Fe 
are solar, but we varied the line widths of O and Fe randomly. For O, we drew a random line width from a 
distribution with an average of once the spatial broadening profile and with a random Gaussian distributed 
variance of 0.25. For Fe, the random width is on average 0.75 with the same variance of 0.25. Using these numbers,
we obtain random O/Fe line width ratios with an average of $\sim$1.3 and a variance of $\sim$0.4. Most of these 
random values will fall in the typically observed range of the O/Fe width ratios between 1-2. For each 
simulated spectrum, we generated a random variety of an asymmetric {\it gdem} multi-temperature model. We drew 
a random number for the sigma parameter for the low-temperature tail with an average of 0.2 and for the 
high-temperature tail of 0.1, both with a variance of 0.05. Each object was assigned a random redshift 
between $z$ 0.01 and 0.1 and an $N_{\mathrm{H}}$ between $10^{19}$ and $10^{22}$ cm$^{-2}$. These numbers 
roughly represent the typical numbers obtained from RGS fits to cluster spectra. The simulated spectra
were subsequently fit with a {\it wdem} model, a single line width, and a free $N_{\mathrm{H}}$. With this
setup, we simulated the effect of bias in the multi-temperature structure, line width, and $N_{\mathrm{H}}$
fitting on a sample of clusters. The simulated biases were random with respect to each other and would
average out for large numbers of simulated clusters.

\begin{figure}
\includegraphics[width=1.0\columnwidth]{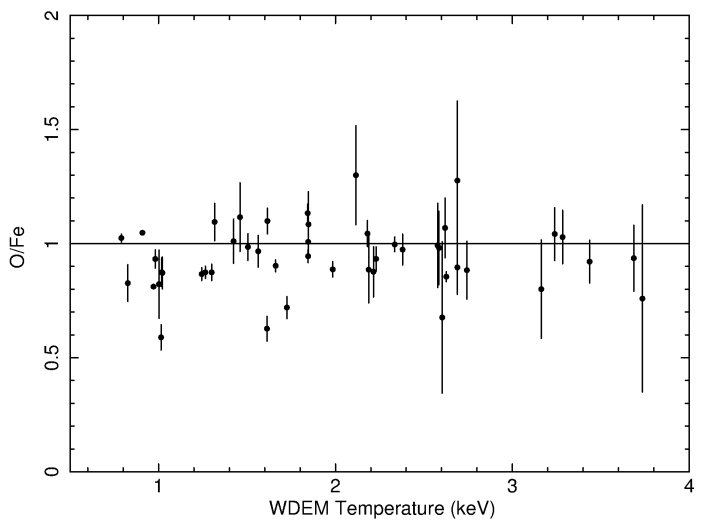}
\caption{O/Fe versus the maximum temperature of a {\it wdem} model ($kT_{\mathrm{max}}$) for a simulated set of 
50 RGS cluster spectra. Three O/Fe ratios are not shown, since the O/Fe ratio was not significantly measured.}
\label{fig:sim-ofe-kt}
\end{figure}

In Fig.~\ref{fig:sim-ofe-kt} we show the O/Fe ratio versus the $kT_{\mathrm{max}}$ temperature from 
a {\it wdem} model fit to the simulated spectra. The O/Fe ratios do not appear to show a significant trend with 
temperature, but there is significant intrinsic scatter in the results. The weighted average of O/Fe in 
the simulated sample is 0.936 $\pm$ 0.005, which is significantly lower than the input value of 1. The
variance of the O/Fe abundances is 0.14. Although this test does not include all possible biases, it shows 
that the biases combined are likely to appear as additional systematic scatter on the O/Fe measurements and
a bias in the average value, similar to what we observe in the CHEERS sample. It is therefore likely that 
a considerable fraction of the scatter in O/Fe observed in the CHEERS sample is due to systematic effects, 
such as inaccurate multi-temperature structure parametrizations or insufficient modeling of line broadening 
effects.

\subsection{An alternative estimation of the total systematic bias}
\label{sec:biastot}

The tests we performed in Section~\ref{sec:fitbias} basically show the sensitivity of our result, the 
O/Fe abundance ratio, to different initial assumptions of the spatial broadening, temperature structure, 
absorption column, and spectral code. This only provides us with a rough indication of what the bias in the
O/Fe ratio may be if our initial assumptions are incorrect. The problem is that we do not know which assumptions 
are right. We try to compare initial assumptions that are generally considered reasonable in the field in
order to obtain a typical bias on the O/Fe ratio. As explained in Section~\ref{sec:biassim}, combining these 
biases into a total bias estimate cannot be calculated exactly. The biases have a direction and magnitude 
that also depends on parameters such as temperature and absorption column, which makes it hard to attach a general 
number on it.

However, we also approached the problem from a different direction because we are able to use constraints from 
our CHEERS sample. Assuming the variation in abundances that we 
measure are only due to systematic uncertainties and that all objects have the same intrinsic abundances. 
Then, following the variance in Table~\ref{tab:abundances}, the typical systematic error for O and O/Fe would be
about 40\% and for Fe nearly 100\%. Because of the limited bandwidth of the RGS and thus the lack of much line-free 
continuum, the absolute abundance of Fe is strongly degenerate with respect to the normalization of the CIE component.
It is therefore better to consider the O/Fe ratio for now, which is not affected by this degeneracy. When we do 
this, the maximum systematic uncertainty would typically be 40\% on O/Fe. It is unlikely, however, that all
local clusters would have exactly the same chemical enrichment history and O/Fe ratio in their cores. The 40\%
can thus be considered to be an upper limit.

When we consider the typical biases in the O/Fe ratio as shown in our simulations, the differences range between 
$\sim10-40\%$, where the 40\% differences only occur in a few extreme cases with large discrepancies in the assumed 
$N_\mathrm{H}$ value. When the biases are combined linearly, the combined bias
typically ranges between $20\%-30\%$ if we exclude some 
exceptional cases. We note that this estimate holds when the latest spectral models (APEC v3.0.1 or SPEXACT v3) are 
used. The older spectral codes show much larger discrepancies. We therefore recommend using the most recent versions
of the spectral codes and being cautious with using literature values for the contribution of molecules to the 
Galactic absorption.

\section{Discussion}

Through the analysis of X-ray spectra of a sample of 44 clusters, groups, and elliptical galaxies, we measured the 
O, Fe, and O/Fe abundance ratios in the cores of these objects. We also estimated the accuracy of our results by testing
the sensitivity of the abundance measurements to our initial assumptions in the data analysis. In this section,
we discuss the astrophysical implications of the O/Fe distribution as a function of temperature and the observed scatter 
in the abundance measurements.

Based on the O/Fe ratio alone, we cannot estimate the fraction of SNIa versus SNcc that enriched the ICM
because the theoretical yields of supernova models vary too much. We need to combine the measured RGS 
abundances with abundances measured using EPIC for a more meaningful comparison with supernova
models. This comparison is reported in \citet{mernier2016b}. 

\subsection{O/Fe ratios and the star formation history of the central BCGs}

The central metal abundance peaks seen in giant elliptical galaxies, groups, and clusters of galaxies 
are thought to be dominated by the metal enrichment from the brightest cluster galaxy (BCG) in the cluster center \citep{degrandi2004}. 
Most BCGs appear red and dead, indicating that they stopped forming stars approximately 10 billion years ago \citep{serra2010}. Today, their 
enrichment is thought to be dominated by SNIa with a long delay time, and to a lesser extent, by metal-rich stellar 
winds. While metal uplift by AGN will only redistribute both SNcc and SNIa products locked up in the stars and the 
gas in the BCG, SNIa explosions are thought to currently dominate the Fe and Ni enrichment. The measured O/Fe ratio is therefore expected 
to decrease with time since the last significant episode of star formation. 

The fact that the O/Fe ratio that we measure does not show a significant trend as a function of temperature 
could indicate that the star formation history of BCGs does not change as a function of cluster mass, which 
is consistent with an earlier conclusion by \citet{degrandi2009}.   
The constant O/Fe may mean that most of the metals were already formed 
before the cluster ICM was assembled around $z\lesssim 2$
\citep{werner2013,simionescu2015} and the contribution from SNIa with long delay times from the BCG is comparatively small.
The constant SNIa/SNcc ratio as a function of cluster radius found by \citet{mernier2017} suggests 
that late enrichment by the BCG either produces similar amounts of SNIa and SNcc, or more likely, that this
late enrichment does, on average, not contribute much to the ICM abundance. The flat radial abundance profiles in the 
very core of clusters could be formed by redistribution of 'old' enrichment products from the starburst period of the BCG 
($z\sim 2-4$) into the ICM by processes such as AGN feedback and sloshing.

\subsection{Intrinsic scatter in the O/Fe abundances}

The measured O/Fe ratios are not consistent with a simple constant value, but there is an additional scatter of 
0.34 in the O/Fe abundance values. In Section~\ref{sec:biassim}
we showed that systematic effects,
such as effects due to uncertainties in the multi-temperature structure or line broadening, contribute to 
the scatter in the O/Fe abundance. Although these biases are largely due to inaccuracies in the spectral 
modeling, a part of the scatter may be due to real variations in cluster cores. It is likely that the true
multi-temperature structure in the cores of clusters and groups depends on the level of AGN activity. The 
chemical enrichment history may also show slight variations between objects. In this section, we explore
possible astrophysical origins for the remaining fraction of the observed scatter.

In Fig.~\ref{fig:hist} we find a weak hint of a tail of objects showing O/Fe ratios that are higher than 
the weighted average, above about 0.9. Although this tail is not formally significant, the distribution may be
broadened and skewed due to intrinsic scatter. This may be partly the result of a late star-forming period in 
some of the the BCGs. This result could be consistent with optical observations of BCGs, which indicate 
that even though most systems stopped forming stars at high rates a long time ago (current star formation 
rates are on the order of $\sim0.1M_{\odot}$ yr$^{-1}$), a few systems show star formation rates of a 
few solar masses per year, very exceptionally up to a few 100 solar masses per year \citep{odea2008,mcdonald2012}. 
In order to test this hypothesis, we compared our O/Fe measurements to estimates of the age of the 
stellar population in elliptical galaxies by \citet{serra2010}. Only eight clusters and groups appear in 
both samples, which limits the comparison. Interestingly, the single stellar population (SSP) equivalent age 
($t_\mathrm{SSP}$) from \citet{serra2010} and our O/Fe measurements show no correlation. A larger
overlapping sample would be necessary to determine whether a significant correlation exists.
Thus, the test of our hypothesis is unfortunately inconclusive.

The scatter in the absolute abundances that we measure appears to be larger for iron than for 
oxygen. Although we should be careful with interpreting absolute abundances with the RGS because of the systematic 
effects, a comparison between the scatter in oxygen relative to
the scatter in iron can be made. The larger scatter in the iron abundance suggests that the object-to-object
variation in SNIa enrichment of the core of the ICM exceeds the SNcc variation, either because of variations 
in the size or age of the stellar populations in the core or variations in the enrichment mechanisms that 
transport the iron from the BCG into the ICM. \citet{ehlert2011} showed that in extreme cases, the metallicity 
peak in the cores of clusters can be mixed into the surrounding gas through violent AGN feedback. Metal 
uplift by AGN-blown bubbles has been observed in many clusters \citep{simionescu2008,simionescu2009,kirkpatrick2009,kirkpatrick2011,kirkpatrick2015}.
Hydrodynamic simulations by \citet{planelles2014} show that AGN feedback is likely to play a major role in 
the radial metal profile. Late-time SNIa enrichment in the BCG may thus be more peaked in one cluster and more 
extended in the other through AGN feedback. This may explain a part of the scatter in O/Fe that we observe in the 
CHEERS sample. However, when we plot the O/Fe abundance ratio and the Fe versus the radio luminosity from \citet{birzan2012}, we
do not see any relation. Since the radio mode activity only shows the recent AGN feedback, an episode
of major feedback in the past could have redistributed the metals in the core a long time ago and therefore
it does not show a correlation between O/Fe and radio luminosity today. Hence, we cannot draw conclusions from the
lack of correlation.

\section{Conclusions}

We have measured the O/Fe abundances in 44 clusters and groups of galaxies with the RGS and summarize our results below.
\begin{itemize}
\item The O/Fe ratio does not show a significant trend as a function of temperature in the 0.6$-$6 keV 
range, which suggests that the enrichment of the ICM does not depend on cluster mass and that
most of the enrichment took place before the ICM was formed. 
\item We estimate that the systematic biases in the O/Fe ratio that are due to the spatial broadening of spectral 
lines in RGS, the multi-temperature structure, and the spectral models are about 20$-$30\%. A thorough 
analysis of the systematic uncertainties proved to be essential to reach this accuracy. 
\item We find a significant scatter in the O/Fe ratio from cluster to cluster. A considerable fraction of
the scatter is most likely due to systematic uncertainties originating from the assumed multi-temperature 
structure, line broadening effects, $N_{\mathrm{H}}$ uncertainties, and uncertainties in spectral models. 
A fraction of the scatter is due to true object-to-object variations in the abundance and
multi-temperature structure, for instance. However, using the current data, we are not able to separate these 
components and estimate the true intrinsic scatter in cluster and group abundances accurately and quantitatively.   
\item Thanks to the high statistics of the observations, we have identified and corrected a bias in the 
old MEKAL code \citep{mewe1985} that caused a bias of about 50\% in the O/Fe ratio. This shows that deep 
observations of samples can help reduce the systematic biases in our analyses. 
\item Before new missions such as XARM and ATHENA fly, it is essential to 
invest in improving atomic line databases and spectral models to reduce their biases to a level that
makes them comparable to the expected level of statistical precision for these instruments.
\end{itemize}

For the systematic biases in the determination of abundances with the RGS, we summarize our findings here.
\begin{itemize}
\item The spatial line broadening bias is about 10\%. This can be reduced by using multiple lpro models to fit 
the widths.
\item The multi-temperature structure bias is about $10-20$\%. Minimizing it requires better observations and 
modeling of cluster cores.
\item The broadband continuum shape bias is typically $\sim$20\%. It can be reduced by fitting the continuum more carefully, 
for instance, free N$_{\mathrm{H}}$ and/or local continuum fitting.
\item The spectral modeling bias is about $10-20$\%. This can only be reduced by improving atomic data and spectral modeling. 

\end{itemize}

\begin{acknowledgements}
We dedicate this paper to our dear colleague and coauthor Yu-Ying Zhang, who passed away on December 11, 2016.
We are deeply grateful for her important scientific contributions and her generous support of the CHEERS project. \\

This work is based on observations obtained with XMM-Newton, an ESA science mission
with instruments and contributions directly funded by ESA Member States and
NASA. Part of the data is obtained through a successful XMM-Newton VLP proposal by
the CHEERS collaboration (proposal ID 072380). 
The Netherlands Institute for Space Research (SRON) is supported financially
by NWO, the Netherlands Organisation for Scientific Research. 
N.W. has been supported by the Lend\"ulet LP2016-11 grant awarded by the Hungarian Academy of Sciences.
C.P. and A.C.F. acknowledge funding through ERC Advanced Grant number 340442.
P.K. thanks Steve Allen and Ondrej Urban for support and hospitality at Stanford University.
Y.Y.Z. acknowledged support by the German BMWi through the Verbundforschung under grant 50~OR~1506.
L.L. acknowledges support from the Transregional Collaborative 
Research Centre TRR33 The Dark Universe and from the HRC contract number NAS8-03060.
H.A. acknowledges support from NWO via a Veni grant.
G.S. acknowledges support from NASA through XMM-Newton grant NNX15AG25G.
T.H.R. and G.S. acknowledge support from the DFG through Heisenberg grant 
RE 1462/5, grant RE 1462/6, and from TRR 33. The authors thank Liyi Gu for helpful 
suggestions regarding the manuscript.
\end{acknowledgements}

\bibliographystyle{aa}
\bibliography{clusters}

\end{document}